\documentclass[a4paper,oneside,12pt]{elsarticle}
\usepackage{amsthm}
\usepackage{amsmath}
\usepackage{graphicx}%
\usepackage{amsfonts}%
\usepackage{amssymb}
 \usepackage{verbatim}
 \usepackage{float}
\usepackage{rotating}
\usepackage{amsmath,amsthm,amssymb}

\newtheorem{defi}{Definition}

\newtheorem{prop}{Proposition}
\newtheorem{rem}{Remark}
\newtheorem{them}{Theorem}

\begin{document}
%\tableofcontents

\title{\huge{\textbf{A Review of Volatility and Option Pricing}}\\
\large{\textbf{by\\
Sovan Mitra}}} \maketitle
\section*{\large{Abstract}}
The literature on volatility modelling and option pricing is a
large and diverse area due to its importance and applications.
This paper provides a review of the most significant volatility
models and option pricing methods, beginning with constant
volatility models up to stochastic volatility. We also survey less
commonly known models e.g. hybrid models. We explain various
volatility types (e.g. realised and implied volatility) and
discuss the empirical properties.
\\\\
\textbf{Key words}: Option pricing, volatility models, risk
neutral valuation, empirical volatility.

\line(1,0){400}
%\newpage
%%%%%%%%%%%%%%%%%%%%%%%%%%%%%%%%%%%%%%%%%%%%%%%%%%%%%%%%%%%%%%%%%%%%%%%%%%%%%%%%%%%%%%%%%%%%%%%%%%%%%%%%%%%%%%%%%%%%%%%%%%%%%%
\begin{comment}
\begin{center}
\begin{tabular}{c}
\\
\\
\\
\\
\\
\\
\\
\\
{\bf \large{``Risk comes from not knowing what you're doing",}}
\\
{\bf \large{Warren Buffett.}}
\\
\end{tabular}
\end{center}

\newpage
\end{comment}
%%%%%%%%%%%%%%%%%%%%%%%%%%%%%%%%%%%%%%%%%%%%%%%%%%%%%%%%%%%%%%%%%%%%%%%%%%%%%%%%%%%%%%%%%%%%%%%%%%%%%%%%%%%%%%%%%%%%%%%%%%%%%%

\section{Introduction and Outline}
This paper provides a review of the most significant volatility
models and their related option pricing models,  where we survey
the development from constant up to stochastic volatility. We
define volatility, the volatility types and study the empirical
characteristics e.g. leverage effect. We discuss the key
attributes of each volatility modelling method, explaining how
they capture theoretical and empirical characteristics of implied
and realised volatility e.g. time scale variance. We also discuss
less commonly known models.

The study of volatility has become a significant area of research
within financial mathematics. Firstly, volatility helps us
understand price dynamics since it is one of the key variables in
a stochastic differential equation governing an asset price.
Secondly, volatility is the only variable in the Black-Scholes
option pricing equation that is unobservable, hence the ability to
model volatility is crucial to option pricing.

Thirdly, volatility is a crucial factor in a wide range of
research areas. For example, contagion effects involve the
``transmission" of volatility from one country to another
\cite{baur2003tcm}. Volatility can explain extreme events as Blake
\cite{FMA} explains that the October 1987 crash could have
resulted from volatility changes.

Finally, volatility has a wide range of industrial applications
from pricing exotic derivatives to asset pricing models
\cite{rebonato2004vac}. Shiller \cite{shiller1989mv} argues the
market's volatility dynamics can be applied to macroeconomic
variables, particularly as the stock market is a well known
leading indicator of the economy. Shiller \cite{UseVol} also
claims volatility can be used as a measure of market efficiency.

Option pricing in itself has become an important research area.
Research interest in options pricing began with the Black-Scholes
option pricing paper \cite{black1973poa};  since then the
derivatives market has grown into a multi-trillion dollar industry
\cite{stout1999lhs}. Options have become important to industry,
particularly as they can be used to hedge out risk. In fact in
many situations it is more attractive to speculators and hedgers
to trade an option rather than an underlying due to the limited
loss. Additionally, option trading can normally be executed on a
far higher level of leverage compared to trading stocks, therefore
offering potentially higher returns for the same initial deposit.

The outline of the paper is as follows. Firstly, we review basic
financial mathematics theory, which is essential for the study of
volatility modelling and option pricing. Next, we introduce the
differing types of volatility and discuss their empirical
behaviour e.g. leverage effect. We then discuss the key models of
volatility and their associated option pricing methods. We finally
end with a conclusion.

\section{Review of Financial Mathematics Theory}\label{Preliminaries chapter}
 This section provides a brief
exposition of standard financial mathematics theory. The
recommended references on this area are Bjork \cite{bjork2004atc},
{\O}ksendal \cite{oksendal2003sde}, Neftci \cite{neftci1996imf}
and Delbaen and Schachermayer \cite{delbaen2006ma}.

\subsection{Stochastic Calculus}
 \subsubsection{Stochastic Differential Equations}
 From the time of Bachelier's work \cite{bachelier1900tst}, the most popular method of modelling asset prices has been using stochastic  differential equations.
 Let $X(t)$ denote a stochastic process (such as a stock price) and we define it as a diffusion if it approximately follows the stochastic difference equation:
\begin{equation}
 X(t+ \Delta t)-X(t)= \mu (t, X(t))\Delta t+\sigma(t,
 X(t))\Delta W(t),
\end{equation}
where:
\begin{itemize}
  \item $\Delta W(t)=W(t+\Delta t)-W(t)$;
  \item $\mu (t, X(t))$  denotes the  drift of
  $X(t)$;
  \item $\sigma(t, X(t))$ denotes the volatility (also known as the
diffusion term).
\end{itemize}
The increment $\Delta W(t)$ is from W, which is called a Wiener
process.
\begin{defi}
A stochastic process W(t) is called a Wiener process if it
satisfies the following conditions:
\begin{enumerate}
\item $W(0)=0$;
\item $W(t)$ has independent increments. In other words $W(u)-W(t)$ and $W(s)-W(r)$ are independent for $r< s\leq
t<u; $
\item $W(t)$ has continuous trajectories;
\item $W(t)-W(s)\sim\mathcal{N}(0,\sqrt{t-s})$ for $s<t$.
\end{enumerate}
\end{defi}
%It can be shown that the trajectories of a Wiener process are
%differentiable nowhere with probability one (~\cite{bjork2004atc}, chapter4).\\
The stochastic process is called a diffusion because the equation
models diffusions in physics. For continuous time modelling we let
$\Delta t \rightarrow0$ and so the stochastic difference equation
becomes:
\begin{eqnarray}\label{SDE general eqn}
 dX(t) &=& \mu (t, X(t))dt+\sigma(t,
 X(t))dW,\\
\mbox{ for }X(0) &=& a,
\end{eqnarray}
where a is a constant. Alternatively, in integral form the
equation becomes:
\begin{equation}\label{SDE integral form}
 X(t)= a+ \int _{0} ^{t}\mu (s, X(s))ds+\int _{0} ^{t} \sigma(s,
 X(s))dW.
\end{equation}
It is worth noting in  equation (\ref{SDE integral form}) that the
first integral is a standard Riemman integral whereas the second
integral is a stochastic integral, that is, an integral with
respect to a Wiener process.

\subsubsection{Stochastic Integrals and Ito's Lemma}
To integrate a stochastic process X(t) we must know the conditions
that guarantee the existence of the stochastic integral. We can
guarantee the integral's existence if the process X(t) belongs to
the class $\mathcal{L}^{2}$. To define the class $\mathcal{L}^{2}$
we must introduce the idea of filtration.
\begin{defi} %\cite{neftci1996imf}p102, shreve p51, etheridge p29
Let $\{\mathcal{F}_{t}\}$ denote the set of information that is
available to the observer. If
\begin{eqnarray}
\mathcal{F}_{s}\subseteq \mathcal{F}_{t}\subseteq \mathcal{F}_{T},
\forall s,t \mbox{ with }s<t<T,
\end{eqnarray}
then the set $\{\mathcal{F}_{t}\},t \in [0,T]$ is called a
filtration.
\end{defi}%\cite{cont2004fmj}p39
For a given stochastic process X(t), increasingly more information
is revealed to an observer as time progresses. Hence at time t=a
(where a is a constant) some information is revealed and this
information is known with certainty at any future time $t>a$. To
keep track of the information flow that is revealed at time t we
introduce the filtration $\mathcal{F}_{t}^{X}$.
\begin{defi} %\cite{bjork2004atc}p39
The notation $\mathcal{F}_{t}^{X}$ denotes the information
generated by process X(t) on the interval [0,t]. If based upon the
observations of the trajectory of X(s) over the interval $s \in
[0,t]$ it is possible to determine if event A has occurred, then
we say A is $\mathcal{F}_{t}^{X}$-measurable and write $A \in
\mathcal{F}_{t}^{X}$. If the stochastic process Z can be
determined based upon the trajectory of X(s) over the interval $s
\in [0,t]$  and $Z(t) \in \mathcal{F}_{t}^{X},\forall t \geq 0$,
then we say Z is adapted to the filtration
$\{\mathcal{F}_{t}^{X}\}_{t\geq 0}$.
\end{defi}
We now define the class $\mathcal{L}^{2}$.
\begin{defi} %\cite{benth2004ots}p36, \cite{bjork2004atc}p40
A stochastic process X(s) belongs to the class
$\mathcal{L}^{2}[a,b]$ if the following conditions are satisfied
\begin{itemize}
  \item X(s) is adapted to the $\mathcal{F}_{t}^{X}$ filtration;
  \item $\int_{a}^{b} E[X^{2}(s)]ds < \infty.$

\end{itemize}
\end{defi}
If we wish to obtain stochastic differential equations we may
require the stochastic version of the classical chain rule of
differentiation. This is known as Ito's lemma.
%Additionally, since working directly with
%stochastic integrals tends to be unoperational, we can deduce
%their integrals by deducing clues from their associated derivatives as in standard calculus.
\begin{them}
(Ito's Lemma) Assume that $X(t)$ is a stochastic process with the
stochastic differential given by
 \begin{equation}
 dX(t)=\mu (t)dt+\sigma (t)dW,
 \end{equation}
 where $\mu$ and $\sigma$ are adapted processes. %cf \cite{benth2004ots}p50
 Let $Z(t)$ be a new process defined
by $Z(t)=f(X,t)$ and f is a twice differentiable function, then Z
has the stochastic differential:
\begin{equation}
df(X(t),t)= \left( \dfrac{\partial f}{\partial t}+
\mu\dfrac{\partial f}{\partial X} +
\dfrac{1}{2}\sigma^{2}\dfrac{\partial^{2} f}{\partial X^{2}}
\right) dt +\sigma\dfrac{\partial f}{\partial X}dW.
\end{equation}
\end{them}

\subsubsection{Geometric Brownian Motion}\label{GBM section}
To model asset prices it is tempting to apply Bachelier's model.
On studying the empirical behaviour of stock prices over time,
Bachelier in his doctoral thesis \cite{bachelier1900tst} first
suggested modelling asset prices $X(t)$ with the following
equation:
\begin{eqnarray}\label{Bachelier eqn}
 dX(t) &=&  \mu Xdt+ \sigma dW,\\%MUST BE THIS VERSION TO BE CORRECT WITH CEV MODEL
\mbox{for }X(0)&=& a,
\end{eqnarray}
where $a,\mu$  and $\sigma$ are constants. However, Bachelier's
equation was not satisfactory;  theoretically stock prices are
always non-negative yet Bachelier's equation allowed negative
stock prices. We would expect percentage returns to be independent
of stock price X(t) yet Bachelier's equation is not.
%DON'T UNDERSTAND SO LEAVE
%Similarly,  % \cite{HULLIntro},p222. if $\sigma$ is interpretted as an investor's uncertainty on
%percentage returns, then this uncertainty should be  at \$5 as at \$50.
%\cite{HULLIntro}.%p223
Samuelson in 1965 \cite{samuelson1965ppa} introduced the Geometric
Brownian motion (GBM) of stock prices, which is the standard model
for stock prices to date:
\begin{eqnarray}
dX/X &=& \mu dt+\sigma dW, \label{GBM eqn}\\
X(t) &=& X(0)exp \left( \left(\mu-\frac{1}{2}\sigma^2
\right)t+\sigma W \right).
\end{eqnarray}
In equation (\ref{GBM eqn}) X(t) is nonnegative with probability
one and if $X(t_{1})=0$ then
\begin{eqnarray}
X(t)=0, \forall t\geq t_{1}.
\end{eqnarray}
This property reflects financial ``bankruptcy" since once X(t)
equals 0 it will remain permanently 0 thereafter. %\cite{fouque2000dfm}p8

If we have a portfolio of n assets
$\mathbf{X}(t)=(X_{1}(t),X_{2}(t),....,X_{n}(t))$, governed by n
independent Wiener processes
$\mathbf{W}=(W_{1}(t),W_{2}(t),....,W_{n}(t))$, we have the model
\cite{etheridge2002cfc}:
\begin{eqnarray}\label{Multistock GBM model}
dX_{i}/X_{i} &=& \mu_{i}dt+\sum_{j=1}^{n}\sigma_{ij}dW_{j}, \mbox{
i=1,..,n and j=1,..,n},\\
X_{i}(t) &=& X_{i}(0)exp
\left(\left(\mu_{i}-\frac{1}{2}\sum_{j=1}^{n}\sigma_{ij}^{2}
\right)t+ \sum_{j=1}^{n} \sigma_{ij}W_{j}\right),
\end{eqnarray}

where
\begin{itemize}
  \item $\sigma_{ij}$ is the volatility matrix
\begin{eqnarray}
\begin{pmatrix}
  \sigma_{11} & \ldots & \sigma_{1n} \\
  \vdots & \ddots & \vdots \\
  \sigma_{n1} & \ldots & \sigma_{nn}
\end{pmatrix};
\end{eqnarray}
  \item the volatility of each stock $X_{i}(t)$ is $\sigma_{i}(t)$ where
\begin{eqnarray}
\sigma_{i}=\sqrt{\sum_{j=1}^{n}\sigma_{ij}^{2}};
\end{eqnarray}
  \item
  instantaneous correlation between $W_{i}$ and $W_{k}$ is given by %\cite{shreve2004scf,p226}
\begin{eqnarray}
corr(dW_{i},dW_{k}) &=&
\rho_{ik}(t)dt\\
&=&
\frac{1}{\sigma_{i}(t)\sigma_{k}(t)}\sum_{j=1}^{n}\sigma_{ij}(t)\sigma_{kj}(t).
\end{eqnarray}
\end{itemize}

\subsection{Black-Scholes Option Pricing}\label{Black-Scholes Option
Pricing section}
%\subsection{replicating portfolio/Black-Scholes} Fouque p12
%1.3-1.3.2 p15, 1.3.4 p15-p18,
%A derivative is: Fouque p8 1.2-p10.\\
The Black-Scholes analysis of European options \cite{black1973poa}
yields a closed form solution to option pricing, only requiring
observable variables (except for volatility). The key insight into
their analysis was to construct a dynamically replicating
portfolio of a European option to value it. Options have become
one of the most important and frequently traded derivatives in
finance. We give a definition of a derivative security as follows \cite{bingham2004rnv}.%p2
\begin{defi}
A derivative security (also known as a contingent claim) is a
financial contract whose value at expiration time T is precisely
determined by the price of an underlying asset at time T.
\end{defi}
As the name implies, an option gives the right (but not the
obligation) to buy or sell an asset at a pre-determined price and
time. A call option gives the right to buy the asset whereas a put
gives the right to sell the asset at a predetermined price. One
can purchase European options, which are options that can only be
exercised at expiration T, or American options, which can be
exercised any time during the life of the option.

The Black-Scholes portfolio replication argument for European call
options begins by considering a market consisting of two assets -a
riskless bond B(t) and a stock X(t) with equations respectively:
\begin{eqnarray}
dB(t) &=& rB(t)dt,\\
 dX(t)/X(t) &=& \mu dt+\sigma dW,
\end{eqnarray}
where $\sigma$ and $\mu$ are constants and r is the risk free rate
of return. Black and Scholes managed to determine the closed form
value of a European call option C on the assumption of no
arbitrage (to be defined in section \ref{Fundamental Theorems}):

\begin{align}\label{Black-Scholes eqn} %agrees with wilmott
C(X(t),t,T,r,\sigma,K)  &= X(t)\Psi(d_1) - Ke^{-r(T-t)}\Psi(d_2),
\\\label{d1 Black-Scholes eqn}
 \mbox{where } d_1&=
\frac{\ln(X(t)/K)+(r+\textstyle\frac{1}{2}\sigma^2)(T-t)}{\sigma
\sqrt{(T-t)}}, \\
d_2&=\frac{\ln(X(t)/K)+(r-\textstyle\frac{1}{2}\sigma^2)(T-t)}{\sigma \sqrt{(T-t)}},\\
&=d_1-\sigma \sqrt{(T-t)}.
\end{align}
In C(X(t),t,T,r,$\sigma$,K) t is the time at which C is being
priced, T is the expiration date, $\Psi(\cdot)$ is the standard
normal cumulative distribution function and K is the strike price.
For a more detailed discussion of the Black-Scholes option pricing
equation the reader is referred to \cite{duffie2001dap}.

\subsection{Risk Neutral Valuation}

\subsubsection{Martingales}\label{martingale definition} In financial mathematics, martingale
processes are important to understanding key concepts, such as the
idea of completeness. We therefore define martingales now.
\begin{defi}
An $\mathcal{F}_{t}-adapted$  process $\mathcal{M}_{t}$ is an
$\mathcal{F}_{t}-martingale$ if
\begin{itemize}
\item $E[\mid \mathcal{M}_{t}\mid ]<\infty, \forall t\geq0; $
\item $E[\mathcal{M}_{s}\mid \mathcal{F}_{t}]=\mathcal{M}_{t}$, for $s\geq t.$
\end{itemize}
If we replace the equality with $\leq$ (or $\geq$) then
$\mathcal{M}_{t}$ is a supermartingale (or submartingale
respectively).
\end{defi}
Whereas a martingale can be considered a fair game, submartingales
and supermartingales  can be considered as favourable and
unfavorable games, respectively, from the gambler's perspective. A
trivial example of a martingale is a constant k;  let $X(t)=k$
then:
\begin{eqnarray}
E[X(t+s)|\mathcal{F}_{t}]=X(t)=k.
\end{eqnarray}
However in financial mathematics we are interested in stochastic
processes since we wish to model asset prices. We therefore define
martingales for stochastic processes.
\begin{prop}
A stochastic process  $X$
 is a martingale if and only if the stochastic differential is of the form
\begin{eqnarray}
dX(t)=f(t)dW,
\end{eqnarray}
where f is any process satisfying the condition:
\begin{eqnarray}
\int_{0}^{t}E[f^{2}(s)]ds\leq \infty.
\end{eqnarray}
\end{prop}
Therefore for stochastic processes to be martingales they must be
``driftless".
%This has important consequences for ensuring
%stochastic processes are martingales.

\subsubsection{Girsanov's Theorem: Change of Measure}
%Prior to the Black-Scholes method [INSERT SECTION/EQN], there
%existed
In addition to the Black-Scholes equation (equation
(\ref{Black-Scholes eqn})), there
exists another method for valuing European options, known as risk neutral valuation. %$http://www.riskglossary.com/link/option_pricing_theory.htm$
Samuelson showed that the price of a call option is equal to its
discounted expected payoff:
%
%
%DON'T \widetilde{r} in eqnarray because cause compile error!!!!!!
%
%
\begin{eqnarray}\label{RNV formula}
C(X(t),K,t,T)=e^{-\tilde{r}(T-t)}E^{\mathbb{P}}[X(T)-K]^{+},
\end{eqnarray}
where $\mathbb{P}$ is the probability measure and $\tilde{r}$ is
the
discount factor for risk. %also see \cite{HULLIntro}p245
The difficulty with this approach is that the discount factor
$\tilde{r}$ and the probability measure $\mathbb{P}$ vary
according to the risk preference of an investor, so were unknown
and typically arbitrarily chosen. For instance a risk neutral (or
risk indifferent) investor does not require an incentive or
disincentive to take on a risky investment, therefore if the
discounted expected value remains constant he will pay the same
price for an investment regardless of the risk.  Furthermore for a
risk neutral investor we have $\tilde{r}$ equals the risk free
rate of return (which is observable), however $\mathbb{P}$ was
unknown at the time Samuelson proposed his option formula.

Cox et al. in \cite{cox1979ops} realised that under a
Black-Scholes model, option pricing is independent of risk
preferences. The investor's risk aversion increases with the
stock's drift $\mu$;  yet $\mu$ does not appear in the
Black-Scholes option pricing
equation (equation (\ref{Black-Scholes eqn})). %\cite{HULLIntro}p245.
By using equation (\ref{RNV formula}), it was shown that all
option prices would give the same option price regardless of the
risk preference chosen provided $\widetilde{r}$ and $\mathbb{P}$
are chosen consistently. Therefore equation  (\ref{RNV formula})
and the Black-Scholes formula would always give the same answer.

To value options as a risk neutral investor, we discount at the
risk free rate and take expectations under the risk neutral
measure.
% Secondly a risk neutral investor expects the expected rate of return on a stock to
%be the same as a riskless bond. %benth p55
%This corresponds to setting $\mu=r$.
To take expectations under the risk neutral measure we need to
change $\mathbb{P}$ in the stochastic differential equation, which
requires Girsanov's Theorem. Girsanov's Theorem tells us how a
stochastic differential equation (SDE) changes as probability
measure $\mathbb{P}$ changes. Essentially, Girsanov's Theorem
tells us a change in $\mathbb{P}$ corresponds to a change in drift
$\mu$ and the rest of the SDE remains unchanged.
%The reader is referred to Neftci \cite{neftci1996imf}Chapter 14 for more
%detail on  change of probability measures.
To explain how we change probability measures, let us define
probability spaces.
\begin{defi} %Netci p79, etheridge p29
The triple $\{\Omega,\mathcal{F},\mathbb{P} \}$ is called a
probability space  where $\Omega$ denotes the sample space, the
set of all possible events,   $\mathcal{F}$ denotes a collection
of subsets of $\Omega$ or events and $\mathbb{P}$ is the
probability measure on $\mathcal{F}$ or events. The quadruple
$\{\Omega,\mathcal{F},\{ \mathcal{F}_{t}\}_{t\geq0},\mathbb{P}\}$
is called a filtered probability space.
\end{defi}
Assume we have the probability space
$\{\Omega,\mathcal{F},\mathbb{P} \}$ then a change of measure from
$\mathbb{P}$ to $\mathbb{Q}$ means we have probability
space $\{\Omega,\mathcal{F},\mathbb{Q} \}$. %BENTH P76
We now state Girsanov's Theorem for change of probability measures
for SDEs.
\begin{defi}%SHREVE BEST REF \cite{neftci1996imf}p290+ BUT ERRORS -CF BJORK, etheridge p98+, SHREVE
(Girsanov's Theorem) Suppose we have a family of  information sets
$\mathcal{F}_{t}$ over a period [0,T] where $T<\infty$. Define
over [0,T] the random process (known as the Doleans exponential)
$\xi_{t}$:
\begin{eqnarray}
\xi_{t}=exp\left\{-\int_{0}^{t}\lambda(u)dW^{\mathbb{P}}(u) -
\frac{1}{2}\int_{0}^{t} \lambda^{2}(u)du \right\},
\end{eqnarray} where $W^{\mathbb{P}}(t)$ is
the Wiener process under probability measure $\mathbb{P}$ and
$\lambda(t)$ is an  $\mathcal{F}_{t}$-measurable process that
satisfies the Novikov condition %SHREVE GIVE SLIGHTLY DIFFERENT BUT MOST OTHERS AGREE WITH BELOW, INCLUDING BJORK
$$E^{\mathbb{P}}\left[exp\left\{\frac{1}{2}\int_{0}^{t}\lambda^{2}(u)du \right\}\right] <\infty, t \in [0,T].$$
%If $\xi_{t}$ is a martingale with respect to the measure $\mathbb{P}$ and information set $\mathcal{F}_{t}$, then -NOT REALLY NECESSARY IF INCLUDE NOVIKOV CONDITION -CF SHREVE, SO BEST TO LEAVE
Then $W^{\mathbb{Q}}$ is a Wiener process with respect to
$\mathcal{F}_{t}$ under probability measure $\mathbb{Q}$, where
$W^{\mathbb{Q}}$ is defined by
\begin{eqnarray}
W^{\mathbb{Q}}(t)=W^{\mathbb{P}}(t)+\int_{0}^{t}\lambda(u)du, t
\in [0,T].
\end{eqnarray}
%The probability measure Q is defined by the Radon-Nikodym
%derivative:
%$$\frac{dQ(\omega)}{dP(\omega)}=\xi$$
%Rearranging gives:
The probability measure $\mathbb{Q}$ is defined by %etheridge p98
\begin{eqnarray}
 \mathbb{Q}[A]=\int_{A}\xi_{t} dP,A \in \mathcal{F}_{t},
\end{eqnarray}
with A being an event in $\mathcal{F}_{t}$.
%$$Q(A)=E^{P}[\mathbf{1}_{A}\xi_{t}] $$
%with A being an event in $\mathcal{F}_{t}$ and indicator function
%$\mathbf{1}_{A}$ of that event.
\end{defi}
Heuristically, Girsanov's Theorem tells us that a change of
measure is a subtraction or addition of an
$\mathcal{F}_{t}$-adapted drift in the Wiener process
\cite{baxter1996fci}:
\begin{eqnarray}
dW^{\mathbb{Q}}=dW^{\mathbb{P}}+\lambda(t)dt.
\end{eqnarray}
Our \textit{discounted} risk neutral process can be expressed as
\begin{eqnarray}
dX/X & = & (\mu-r)dt + \sigma dW^{\mathbb{P}},\\
     & = & (\mu-r)dt + \sigma(dW^{\mathbb{Q}}-\lambda(t)dt).
%     & = & \sigma dW^{Q}, \mbox{ for } \mu=r
\end{eqnarray}
%Therefore under a discounted risk neutral process since $\mu=r$
%the drift is cancelled out. Therefore we obtain a martingale as
%defined in [INSERT NO.] under measure Q. In other words,
If we choose
\begin{eqnarray}
\lambda(t)=(\mu -r)/\sigma,
\end{eqnarray}
we obtain equation
\begin{eqnarray}
dX/X=\sigma dW^{\mathbb{Q}},
\end{eqnarray}
which is a martingale (see section \ref{martingale definition})
since $\lambda(t)$ cancels the drift. This choice of $\lambda(t)$
is also known as the market price of risk and the Sharpe ratio
(discovered by Sharpe \cite{sharpe1966mfp}). Note that only one
choice of $\lambda(t)$ in this SDE gives a martingale (or
alternatively eliminates the drift) hence $\mathbb{Q}$ is known as
the unique equivalent martingale measure. The equivalent
martingale measure is also known as the risk neutral measure and
pricing under this measure is known as risk neutral valuation.
%leave because irrelevant
%It is interesting to note that a change of measure
%can be introduced to improve importance sampling (see \cite{steele2001sca}).%steele p213+

\subsubsection{Multidimensional or Multifactor Girsanov's Theorem}%: Multidimensional Change of measure
To change measures for multidimensional SDEs we require the
multidimensional Girsanov's Theorem, which is very similar to the
one dimensional version.% for we can consider the 1D case as a
%special case of the multidimensional version.
\begin{defi}%ehteridge p166, PRETTY SURE BELOW CORRECT BUT CREATED FROM NO. DIFFERENT BOOKS
(Multidimensional Girsanov's Theorem) Suppose we have a family of
information sets $\mathcal{F}_{t}$. Let
$\theta(t)=(\lambda_{1}(t),\lambda_{2}(t),...,\lambda_{n}(t))$ be
an n-dimensional process that is $\mathcal{F}_{t}$-measurable %BENTH P85
 and satisfies the Novikov condition:
\begin{eqnarray}
E^{\mathbb{P}}\left[exp\left\{\frac{1}{2}\int_{0}^{t}
\sum_{i=1}^{n} \lambda^{2}_{i}(u)du \right\} \right] <\infty.
\end{eqnarray}
We define the random process  $\xi_{t}$:
\begin{eqnarray}
\xi_{t}=exp\left\{\sum_{i=1}^{n}
\left(-\int_{0}^{t}\lambda_{i}(u)dW^{\mathbb{P}}_{i}(u) -
\frac{1}{2} \int_{0}^{t} \lambda^{2}_{i}(u)du \right) \right\},
\end{eqnarray}
where  $dW^{\mathbb{P}}_{i}$ for i=1,...,n is an n-dimensional
Wiener process under probability measure $\mathbb{P}$. Then under
the measure $\mathbb{Q}$, $W^{\mathbb{Q}}_{i}$ is a
multidimensional Wiener process defined by:
\begin{eqnarray}
W_{i}^{\mathbb{Q}}=W_{i}^{\mathbb{P}}+\int_{0}^{t}\lambda_{i}(u)du,
\mbox{for }i=1,2,...n.
\end{eqnarray}
The probability measure $\mathbb{Q}$ is defined by
\begin{eqnarray}
 \mathbb{Q}[A]=\int_{A}\xi_{t}dP, A \in \mathcal{F}_{t}.
\end{eqnarray}
\end{defi}
%\cite{benth2004ots,84}
We wish to obtain an equivalent martingale measure $\mathbb{Q}$
related to the \textit{vector} valued  Wiener process
$\mathbf{W}=(W_{1},W_{2},...,W_{n})$. Therefore \textit{each}
stock in the multiple stock model must be a martingale with
respect to $\mathbb{Q}$. Since changing measure means changing
drift we have a vector
$\theta(t)=(\lambda_{1}(t),....,\lambda_{n}(t))$ to give an
n-dimensional $\mathbb{Q}$-Wiener process
\begin{eqnarray}
W_{i}^{\mathbb{Q}}=W_{i}^{\mathbb{P}} +
\int_{0}^{t}\lambda_{i}(s)ds.
\end{eqnarray}
In the one dimensional case we obtained a martingale by setting
$\lambda=\frac{\mu-r}{\sigma}$, in the multiple stock model we
apply a similar argument but in
every dimension: %\cite{Bax,p187-8}
\begin{eqnarray}
\lambda_{i}(t)=\dfrac{(\mu_{i}-r)}{\left(\sum_{j=1}^{n}\sigma_{ij}(t)
\right)},\mbox{ for } i=1,..,m.
\end{eqnarray}
When the volatility matrix is invertible then unique $\lambda_{i}$
solutions exist, admitting a unique risk neutral process. %\cite{baxter1996fci}p187-8, BENTH P87,
Therefore the discounted risk neutral process for the multiple
stock model is:
\begin{eqnarray}
dX_{i}/X_{i}=\sum_{j=1}^{n}\sigma_{ij}d{W}_{i}^{\mathbb{Q}}.
\end{eqnarray}\label{multiple stock discounted RN process}

\subsection{Fundamental Theorems of Finance}\label{Fundamental
Theorems} Before stating the fundamental theorems of finance, we
must define arbitrage and a self-financing portfolio. A portfolio
$V(t)$ is self-financing if the change in value of the portfolio
is a result of the assets' changing values and no external
withdrawals or additions to the portfolio are made. For instance a
portfolio V(t) consisting of $k_{1}$ shares of $X(t)$ and $k_{2}$
bonds of $B(t)$ is a self-financing portfolio if: %p81 baxter
\begin{eqnarray}
dV(t)=k_{1}dX(t) + k_{2}dB(t).
\end{eqnarray}
%In reality there can only exist 1 riskless rate of return, leading
%to the law of one price.
We now define arbitrage.
\begin{defi}%\cite{bjork2004atc}p92, benth p59:
An arbitrage possibility in a financial market is a self-financed
portfolio $V(t)$ such that:
\begin{itemize}
  \item $V(0) \leq 0; $
  \item $V(T)\geq 0$ almost surely and
  \item $E[V(T)]\geq0.$
\end{itemize}
\end{defi}
In words arbitrage is an event where it is possible to make a
profit without the possibility of incurring a loss. We
 say we have a ``fair price" when an asset is free from arbitrage. %Neftci p12, benth p56-9
We are now in a position to state the two key theorems of
financial mathematics.
\begin{them} (First Fundamental Theorem of Finance) The market is
arbitrage free if and only if there exists an equivalent
martingale measure.
\end{them}
\begin{them}%bjork p146
(Second Fundamental Theorem of Finance) Assume that the market is
arbitrage free. The market is then complete if and only if there
exists a unique equivalent martingale measure.
\end{them}
In a no arbitrage market that is incomplete we have a variety of
no arbitrage option prices since we have a variety of martingale measures. %FOUQUE p46
Each martingale measure will produce a different price, in
general.

Completeness and arbitrage are separate properties and we can
illustrate this following the heuristic argument of Bj\"{o}rk
\cite{bjork2004atc}. Let %8.3
$\varpi_{2}$ denote the number of independent sources of
randomness and let $\varpi_{1}$ denote the number of tradable
assets. Now the market is arbitrage free if $\varpi_{1}\leq
\varpi_{2}$ and complete if $\varpi_{1}\geq \varpi_{2}$. Hence the
market is complete and arbitrage free if $\varpi_{1}=\varpi_{2}$.

For no arbitrage we must have $\varpi_{1}\leq \varpi_{2}$ since
each new tradable asset  gives an opportunity to create an
arbitrage portfolio. For example, let $\varpi_{2}$=1 and
$\varpi_{1}$=2 and each asset follows GBM that are identical
(including the same Wiener process since $\varpi_{2}$=1) but have
different drifts:
\begin{eqnarray}
dX_{i}/X_{i}=\mu_{i}dt+\sigma dW.
\end{eqnarray}
Then one can produce a riskless profit by shorting the lower drift
asset $X_{1}$ and using the proceeds to purchase the higher drift
asset $X_{2}$. For completeness we must have $\varpi_{1}\geq
\varpi_{2}$ to enable us to trade or replicate every possible
claim.
%%%%%%%%%%%%%%%%%%%%%%%%%%%%%%%%%%%%%%%%%%%%%%%%%%%%%%%%%%%%%%%%%%%%%%%%%%%%%%%%%%%%%%%%%%%%%%%%%%%%%%%%%%
\begin{comment}
NOT PROVIDE EXTRA INFO SO NOT INCLUDE

We can formalise these ideas of arbitrage and completeness:
\begin{prop} %\cite{benth2004ots}p87
Assume the stocks are modelled as in equation (\ref{GBM eqn}).
Then the market is:
\begin{itemize}
  \item arbitrage free if and only if $\varpi_{1} \leq \varpi_{2}; $
  \item complete if and only if $\varpi_{2} \geq \varpi_{1}; $
  \item complete and arbitrage free if and only if $\varpi_{1}=\varpi_{2}.$
\end{itemize}
\end{prop}
\end{comment}
%%%%%%%%%%%%%%%%%%%%%%%%%%%%%%%%%%%%%%%%%%%%%%%%%%%%%%%%%%%%%%%%%%%%%%%%%%%%%%%%%%%%%%%%%%%%%%%%%%%%%%%%%%

The no arbitrage definition enables us to state the put-call
parity relationship for European options, which does not require
any assumptions other than the market
is arbitrage free. %bjork p123,Hull p173.Kwok p19
\begin{prop}(Put-Call Parity)\\
Assume the market is arbitrage free. If a European call C(t,X) and
a European put P(t,X) with the same underlying asset X, strike K
and expiration T exist then we have the put-call parity relation:
\begin{eqnarray}\label{put-call eqn}
P(t,X)=Ke^{-r(T-t)}+C(t,X)-X(t)+D,
\end{eqnarray}
where D is the cash dividend received from the underlying stock
during the life of the option.
%This relation means that put P(t,X) can be replicated by a
%portfolio containing a zero coupon riskless bond with face value K
%at time t, a call option C(t,X) and a short position in underlier X.
\end{prop}

\subsection{Feynman-Kac Stochastic Representation Formula}\label{Feynman-Kac Stochastic Representation
Formula}
%http://en.wikipedia.org/wiki/Feynman_Kac_Formula
The Feynman-Kac theorem provides a link between the partial
differential equation of  a diffusion process and its expectation.
This is useful because we can solve important analytical problems
either by taking expectations under a risk neutral measure or by
solving the partial differential equation.
\begin{them}(Feynman-Kac Theorem)

Let X(t) satisfy the equation
\begin{eqnarray}
 dX(t) = \mu(t,X(t)) dt + \sigma(t,X(t))dW,
\end{eqnarray}
with initial value at initial time s
\begin{eqnarray}
X(s)=x,
%although at any point X(t) takes on a value of a constant, we want to
%differntiate X(t) at a specific point,x. So X(t) is still variable
%we want a specific value of X(t), that is x.
%
\end{eqnarray}
and let $f(s,x) \in \mathcal{L}^{2}$  be a function  which
satisifies
\begin{eqnarray}
\int_{0}^{T}E \left[\left(\sigma(t,X(t))\dfrac{\partial
f}{\partial x}(t,X(t))\right)^{2} \right]dt < \infty
\end{eqnarray}
 and boundary condition
\begin{eqnarray}
f(T,x)=h(x).
\end{eqnarray}
If the function f(s,x) is a solution to the boundary value problem
\begin{eqnarray}
\dfrac{\partial f}{\partial s}+
\dfrac{1}{2}\sigma^{2}(s,x)\dfrac{\partial^{2} f}{\partial x^{2}}+
\mu(s,x)\dfrac{\partial f}{\partial x}-rf(s,x)=0,
\end{eqnarray}
then f has the representation:
\begin{equation}
  f(s,x)=e^{-r(T-s)}E[h(X(T))|X(s)=x].
\end{equation}
\end{them}
%IF necessary goto etheridge p170 multidimensional F-K formula but not fully understand.
The Feyman-Kac formula can be extended to the case of
$\mathbf{W}=(W_{1},W_{2})$ where $W_{1},W_{2}$ are independent
Wiener processes and we have two stochastic
differentials in $X_{1}$ and $X_{2}$. Following Shreve \cite{shreve2004scf}, %P277
let the stochastic processes follow
\begin{align}
 dX_{1}(t) &=& \mu_{1}(t,X_{1}(t),X_{2}(t)) dt +
 \sigma_{11}(t,X_{1}(t),X_{2}(t))dW_{1}
 + \sigma_{12}(t,X_{1}(t),X_{2}(t))dW_{2}, \label{F-K 2D1}\\
dX_{2}(t) &=& \mu_{2}(t,X_{1}(t),X_{2}(t)) dt+
 \sigma_{21}(t,X_{1}(t),X_{2}(t))dW_{1}
 +  \sigma_{22}(t,X_{1}(t),X_{2}(t))dW_{2}, \label{F-K 2D2}
\end{align}
with initial values at initial time s
\begin{eqnarray}
X_{1}(s) &=& x_{1},\\
X_{2}(s) &=& x_{2},
\end{eqnarray}
with conditions
\begin{eqnarray*}
&\int_{0}^{T}& E[(\sigma_{11}(t,X_{1}(t),X_{2}(t))\dfrac{\partial
f}{\partial
x_{1}}(t,X_{1}(t),X_{2}(t)))^{2}\\
&+& (\sigma_{12}(t,X_{1}(t),X_{2}(t))\dfrac{\partial f}{\partial
x_{1}}(t,X_{1}(t),X_{2}(t)))^{2}]dt < \infty,\\
&\int_{0}^{T}& E[(\sigma_{21}(t,X_{1}(t),X_{2}(t))\dfrac{\partial
f}{\partial
x_{2}}(t,X_{1}(t),X_{2}(t)))^{2}\\
&+& (\sigma_{22}(t,X_{1}(t),X_{2}(t))\dfrac{\partial f}{\partial
x_{2}}(t,X_{1}(t),X_{2}(t)))^{2}]dt < \infty,
\end{eqnarray*}
with boundary condition
\begin{eqnarray}
f(T,x_{1},x_{2})=h(x_{1},x_{2})
\end{eqnarray}
and  satisfy:
\begin{eqnarray}
f(s,x_{1},x_{2})=e^{-r(T-s)}E[h(X_{1}(T),X_{2}(T))|X_{1}(s)=x_{1},X_{2}(s)=x_{2}].
\end{eqnarray}
We then have the associated partial differential equation:
\begin{eqnarray*}
0 &=& \dfrac{\partial f}{\partial s}+
\mu_{1}(s,x_{1},x_{2})\dfrac{\partial f}{\partial
x_{1}}+\mu_{2}(s,x_{1},x_{2})\dfrac{\partial f}{\partial x_{2}}\\
&+& \dfrac{1}{2}\left(\sigma^{2}_{11}(s,x_{1},x_{2}) +
\sigma^{2}_{12}(s,x_{1},x_{2}) \right)\dfrac{\partial^{2}
f}{\partial
x_{1}^{2}}\\
 &+&
\left(\sigma_{11}(s,x_{1},x_{2})\sigma_{21}(s,x_{1},x_{2})+\sigma_{12}(s,x_{1},x_{2})\sigma_{22}(s,x_{1},x_{2})
\right)\dfrac{\partial^{2}
f}{\partial x_{1} \partial x_{2}} \\
&+& \dfrac{1}{2}\left(\sigma^{2}_{21}(s,x_{1},x_{2}) +
\sigma^{2}_{22}(s,x_{1},x_{2})\right)\dfrac{\partial^{2}
f}{\partial x_{2}^{2}}-rf(s,x_{1},x_{2}).
\end{eqnarray*}

\section{An Introduction to Volatility} %taken \cite{bjork2004atc}p105-6

\subsection{Different Types of Volatility}%USEFUL BUT BEST TO LEAVE OUT
Wilmott in \cite{wilmott2001pwi} distinguishes between four
different types of volatility, although in practise little
distinction is made between them. The four types of volatility
are:
\begin{itemize}
  \item volatility $\sigma$ in equation (\ref{SDE general eqn}). It is
   a measure of  randomness in asset return and since it exists at
   each moment in time it has no ``timescale" associated with it.
   For example volatility can be 20\%,5\%,60\%.
  \item historic volatility (also known as realised volatility):
  this is a measure of volatility using past empirical price data
  and will be explained further in section \ref{Historic and Implied
Volatility}.
  \item implied volatility: this is the volatility associated with
  empirical option prices and will be explained further in section
  \ref{Historic and Implied
Volatility}.
 \item forward volatility: this is the volatility obtained from
 some forward instrument.
\end{itemize}
We note that all four types of volatility \textit{in theory}
should not differ since they all refer to the same variable
$\sigma$, however \textit{in practise} they may be different. For
example some researchers believe actual volatility and implied
volatility are 2 separate variables and treat them differently
(see Schonbucher's model in section \ref{Other Volatility Models
section} for instance).

\subsection{Historic and Implied Volatility}\label{Historic and
Implied Volatility} In theory volatility should not depend on the
method of measurement. However, in practice this is not the case.
Volatility $\sigma$ can be empirically measured by two methods:
historic volatility and implied volatility. Historic volatility is
calculated from empirical (and therefore discrete) stock price
data $X(t_{0}),....,X(t_{n})$ where $\Delta t=t_{i}-t_{i-1}$
denotes the chosen sampling interval. To estimate historic
volatility $\widehat{\sigma}$  we calculate the standard deviation
of an asset's continuously compounded return per unit time:
%One may wish to note that the above formula implies
%$$X(t_{i})=X(t_{i-1})e^{\xi_{i}}$$
%Therefore one could interpret $\xi_{i}$ as the the continuously
%compounded return over $\Delta t$ ?????
\begin{eqnarray}
\widehat{\sigma}= \frac{\sqrt{V_{\chi}}}{\sqrt{\Delta t}},
\label{empirical vol eqn}
\end{eqnarray}
where
\begin{itemize}
  \item  $V_{\chi}$ is  the sample variance
\begin{eqnarray}
V_{\chi}=\frac{1}{n-1} \sum_{i=1}^{n}(\chi_{i}-\overline{\chi}).
\end{eqnarray}
Note that sample variance contains n-1 in the denominator, whereas
variance of a theoretical distribution contains n;
\item
\begin{eqnarray}
 \chi_{i}=ln \left( \dfrac{X(t_{i})}{X(t_{i-1})} \right);
\end{eqnarray}
  \item $\overline{\chi}$ is the sample mean
\begin{eqnarray}
\overline{\chi}=\dfrac{1}{n}\sum_{i=1}^{n}\chi_{i}.
\end{eqnarray}
\end{itemize}

In contrast to historic volatility we obtain implied volatility
values from empirical options data. Using Black-Scholes option
pricing, call options C are a function of C(X,t,T,r,$\sigma$,K),
with all the independent variables observable except $\sigma$.
Since the quoted option price $C^{obs}$ is observable, using the
Black-Scholes formula we can therefore calculate or imply the
volatility that is consistent with the quoted options prices and
observed variables. We can therefore define implied volatility
$\mathcal{I}$ by:
\begin{eqnarray}
C_{BS}(X,t,T,r,\mathcal{I},K)=C^{obs}, %\cite{fouque2000dfm}p34
\end{eqnarray}
where $C_{BS}$ is the option price calculated by the Black-Scholes
equation (equation (\ref{Black-Scholes eqn})).
%It is currently a matter of research as to whether implied or
%historic volatility  provides better future volatility forecasts.
Implied volatility surfaces are graphs plotting  $\mathcal{I}$ for
each call option's  strike K and expiration T. Theoretically
options whose underlying is governed by GBM should have a flat
implied volatility surface, since volatility is a constant;
however in practise the implied volatility surface is not flat and
$\mathcal{I}$ varies with K and T.

Implied volatility plotted against strike prices from empirical
data tends to vary in a ``u-shaped" relationship, known as the
\emph{volatility smile}, with the lowest value normally at X=K
(called ``at the money" options).  The opposite graph shape to a
volatility smile is known as a volatility frown due to its shape.
The smile curve has become a prominent feature since the 1987
October crash (see for instance \cite{bates2000pcf} and
\cite{christensen1998rbi}).

Various explanations have been proposed to account for volatility
smiles. Firstly, it has been suggested that options are priced
containing information about future short term volatility that is
not already contained within past price information (see for
instance \cite{jorion1995pvf}). Secondly, implied volatility is
influenced more by market sentiment rather than by pure
fundamentals, for instance the VIX (a weighted average of implied
volatilities from S\&P 500 index options) is used as a gauge of
market sentiment \cite{simon2001spf}.

Thirdly, the transaction costs involved in trading options is
significantly higher and more complicated compared to their
associated underlyings, creating volatility smiles (see for
instance \cite{pena1999wsd}). Finally, Jarrow and Turnbull
\cite{jarrow1996ds} discuss non-simultaneous price observation;
since option and stock prices are from two different financial
markets, there will always be observation time differences. Such
time differences can cause substantial differences in the
estimated implied volatility.

\subsection{Empirical Characteristics of Volatility}\label{Empirical Characteristics of
Volatility} Volatility has been observed to exhibit consistently
some empirical characteristics, %. Such characteristics that are
%common to a wide class of assets are called stylised facts,
which we will now discuss. %(see \cite{cont2001epa}).
Firstly,  one of the most well known empirical characteristics of
volatility is the leverage effect, as first proposed by Black
\cite{black1976ssm}. Black observed that volatility is negatively
correlated with stock price and accounted for this through the
concept of leverage. Leverage (also known as the gearing or debt
to equity ratio) is defined by:
\begin{center}
Leverage=$\dfrac{\iota}{MKT}$,
\end{center}
where $\iota$ is the company's total debt and MKT is the market
capitalisation (number of shares $\times$ share price). As the
share price drops the company becomes riskier, since a greater
percentage of the company is debt financed, hence increasing
volatility.

Black argued leverage could not entirely explain volatility since
companies with little or no debt still exhibit high volatility.
Other explanations of volatility's negative corelation with stock
price include portfolio rebalancing, where investors are forced to
liquidate their assets if they fall below a threshold price.
Alternatively, threshold prices may act as triggers for sales when
interpretted in terms of prospect theory, as proposed by Kahneman
and Tversky \cite{kahneman1979pta}. Also there exists ownership
concentration, where an investor owns a substantial percentage of
a company's stock and sells all his holdings at once (see for
instance \cite{cao2008ead}). In both cases a large volume of
selling pushes prices down further, increasing volatility.

Secondly, volatility (and return distributions) show dependency on
the time scale $\Delta t$ chosen to measure it, as defined in
equation (\ref{empirical vol eqn}). The return distribution
becomes increasingly more Gaussian as $\Delta t$ increases
\cite{cont2001epa}, known as ``Gaussian aggregation", yet under
GBM volatility is theoretically scale invariant. Such empirical
observations have motivated researchers to seek models that
exhibit scale variation
 and is one of the benefits of mean reverting stochastic
volatility models.

Thirdly, Mandelbrot \cite{MANDVar} and Fama \cite{fama1965bsm}
were the first to observe volatility clusters (positively
autocorrelates) with time;  large (small) price changes tend to
follow large (small) price changes. Such observations motivated
GARCH and stochastic volatility models (see section
\ref{stochastic volatility models}) and for this reason volatility
clustering is sometimes known as the
``GARCH effect" \cite{cont2004vcf}. %in  \cite{cont2004vcf} vol clustering is autocorrelation effect
The autocorrelation is significant over time scales $\Delta t$ of
days and weeks but insignificant over longer time scales. This is
because the autocorrelation strength decays following a power law
with $\Delta t$ increasing.

The autocorrelation's slow power law of decay has been cited as
evidence of the existence of long memory in volatility
\cite{cont2001epa} and empirical evidence can be found in
\cite{liu1999spv}. A stationary stochastic process X(t) has long
memory  if its covariance $\upsilon(\tau)=cov(X(t),X(t+ \tau))$
follows the power law of decay \cite{tan2006lmv}
\begin{eqnarray}
\upsilon(\tau)=\tau^{d},\mbox{ for }-1<d<0,
\end{eqnarray}
and
\begin{eqnarray}
\sum_{\tau=0}^{\infty}\upsilon(\tau)=\infty.
\end{eqnarray}

Fourthly, volatility tends to be correlated with significant
financial or economic variables. For instance Schwert
\cite{schwert1989dsv} empirically shows that excessive volatility %\cite{schwert1989dsv}
increases during financial crises; during the Great Depression
(1929-39) stock volatility increased by a factor two or three
times its typical values.
%Officer 1973 "the variability of market....",
% Fama and Schwert 1977 observed high volatility was highly correlated with
%macroeconomic variables e.g. interest rates, inflation,

Finally, in commodities and other asset classes seasonal effects
and calendar time play a significant role in return and volatility
distributions. %NOTE: applies to returns, not necessarily vol\\
%For instance the old trading saying "sell in May...return in
%September" reflects the empirical observations stock prices tend
%to decline during May-September. The January effect relates to the
%fact stock prices tend to rise in this month.
%the calendar effects include, the Day of the Week Effect, Weekend
%Effect (French in "Stock returns and the weekend effect", Journal
%of financialEconomics, Volume 8, Issue 1, March 1980, Pages 55-69
%Kenneth R. French.
For instance,  French \cite{french1980sra} noticed stock returns
measured over the weekend on average are more negative than any
other day of the week. In the past decade the increasing
availability of intraday data has enabled researchers to show
times of the day impact volatility, therefore complicating
intraday volatility estimation \cite{genccay2001dis}.

Other volatility empirical characteristics include mean reversion,
where volatility tends to revert around some long term value
\cite{stein1991spd}. This has motivated mean reverting volatility
models (see section \ref{stochastic volatility models} for more
detail). Volatility tends to be correlated with high trading
volume \cite{scheinkman2003oas} and company specific news (e.g.
earnings
announcements) \cite{rossi1996msm}. Finally % Rossi bk p65 gives paper refs
 Shiller  claims market volatility can
be explained in terms of psychological factors such as fashion, herding %(p9)
 effects %(p13-4),
and overreaction to new information \cite{shiller1989mv}.
% and how opinions "diffuse".
%Shiller mentions opinion diffusion has been modelled using the
%mathematics of epidemics (p14-5).
Shiller in \cite{shiller1989mv} also proposes a stock price model
taking into account psychological factors.
%adding to the dividend
%discount model the present value of the expected future demand of
%ordinary investors (p50).
%Large literature exists on how heterogeneous beliefs affect
%volatility, see "Overconfidence and Speculative Bubbles,Scheinkman and Xiong, causing overvaluation and a driving force for trading
%itself. Also supply and demand imbalance: more natural to go long or adopt a buy-and-hold strategy rather than shorting. This in
%turn creates more demand for downward protection.???????????options?

\section{A Review of Volatility and Option Pricing Models}\label{Volatility Modelling Literature Review chapter}
As described in section \ref{GBM section}, Bachelier
\cite{bachelier1900tst} proposed a model for stock prices, with
constant volatility (see equation (\ref{Bachelier eqn})).
Bachelier reasoned that investing was theoretically a ``fair game"
in the sense that statistically one could neither profit nor lose
from it. Hence Bachelier included the Wiener process to
incorporate the random nature of stock prices. Osborne
\cite{osborne1959bms} conducted empirical work supporting
Bachelier's model. Samuelson \cite{samuelson1965ppa} continued the
constant volatility model under the Geometric Brownian Motion
stock price model (see equation (\ref{GBM eqn})) on the basis of
economic justifications.

Over time, empirical data and theoretical arguments found constant
volatility to be inconsistent with the observed market behaviour
(such as the leverage effect, as discussed in section
\ref{Empirical Characteristics of Volatility}). A plot of the
empirical daily volatility of the S\&P 500 index clearly shows
volatility is far from constant. This consequently led to the
development of dynamic volatility modelling. Volatility modelling
may be classified into four categories:
\begin{enumerate}
\item Constant volatility $\sigma$;
\item Time dependent volatility $\sigma(t)$;
\item Local volatility: volatility dependent on the stock price
$\sigma(X(t))$;
\item Stochastic volatility: volatility driven by an \textit{additional} random process
$\sigma(\omega)$.
\end{enumerate}
These models will now be discussed in some detail in the
subsequent sections.

It is worth mentioning that some models do not use Wiener
processes to capture price movements. For example, Madan and
Seneta \cite{madan1990vgv} propose a Variance-Gamma model where
price movements are purely governed by a discontinuous jump
process. Carr et al. discuss various jump processes in
\cite{carr2002fsa} such as the pure jump process of Cox and Ross
\cite{cox1976ValCEV}. The advantages of pure jump models are:
firstly, they realistically reflect the fact that trading occurs
discontinuously, secondly price jumps are consistent with
information releases according to Fama's ``Efficient Market
Hypothesis" \cite{fama1965bsm} and finally, stock price processes
jump even on an intraday time scale \cite{cont2004fmj}.%p3

%Barndorff-Nielsen and Shepherd \cite{barndorffnielsen2001ngo} use
%Levy processes (see section [INSERT NO]),

\subsection{Time Dependent Volatility Model}\label{Time Dependent Volatility
Model}
%\subsection{General Time Dependent Models}
%$http://www.xplore-stat.de/tutorials/stfhtmlnode52.html$\\
As mentioned in section \ref{Empirical Characteristics of
Volatility}, it was empirically observed that implied volatility
varied with an option's expiration date. Consequently, a straight
forward extension proposed to the constant volatility model was
time dependent
volatility modelling \cite{wilmott1998dta}:%p289
 \begin{eqnarray}
dX/X=\mu dt+\sigma(t)dW.
\end{eqnarray}
Merton \cite{merton1973tro} was the first to propose a formula for
pricing options under time dependent volatility.
The option price associated with X is still calculated by the standard Black-Scholes %wilmott
formula (equation (\ref{Black-Scholes eqn})) except we set
$\sigma={\sigma}_{c}$ where:
%NOTE: \sigma_{c}=\overline{\sigma^{2}} in Fouque p51
%\cite{fouque2000dfm,p39}:
\begin{eqnarray}\label{sigma_c}
\sigma_{c}=\sqrt{\dfrac{1}{T-t}\int^{T}_{t}\sigma^2(\tau)d\tau},
\end{eqnarray}
i.e. $d_{1}$ and $d_{2}$ in the Black-Scholes equation become:
\begin{eqnarray}
d_{1} &=& \dfrac{log \left(\dfrac{X}{K}\right) +\mu(T-t)+\dfrac{1}{2}\int^{T}_{t} \sigma^2(\tau) d\tau}{\sqrt{\int^{T}_{t}\sigma^2(\tau)d\tau}},\\
d_{2} &=& \dfrac{log \left(\dfrac{X}{K}\right)
+\mu(T-t)-\dfrac{1}{2}\int^{T}_{t} \sigma^2(\tau)
d\tau}{\sqrt{\int^{T}_{t}\sigma^2(\tau)d\tau}}.
\end{eqnarray}
The equation (\ref{sigma_c}) converts $\sigma(t)$ to its constant
volatility equivalent $\sigma_{c}$ over time period t to T. The
distribution of X(t) is given by:% \cite{fouque2000dfm,p39}
\begin{eqnarray}
log(X_{T}/X_{t})\sim \mathcal{N}\left((\mu-
\frac{1}{2}\sigma_{c}^{2})(T-t),\sigma_{c}^{2}(T-t)\right).
\end{eqnarray}
Note that the constant volatility $\sigma_{c}$ changes in value as
t and T change. This property enables time dependent volatility to
account for empirically observed implied volatilities increasing
with time (for a given strike).

\subsection{Local Volatility Models}
\subsubsection{Definition and Characteristics}
In explaining the empirical characteristics of volatility, a time
dependent volatility model was found to be insufficient. For
instance, time dependent volatility  did not explain the
volatility smile, nor the leverage effect (see section
\ref{Empirical Characteristics of Volatility}), since volatility
cannot vary with price. Therefore volatility as a function of
price (and optionally time) was proposed, that is \textit{local
volatility} is
$\sigma=f(X,t)$: %(also known as restricted volatility and forward (forward) volatility \cite{wilmott1998dta,p292}):
\begin{eqnarray}
dX/X=\mu dt+\sigma(X,t)dW.
\end{eqnarray}
The term ``local" arises from knowing volatility with certainty
``locally" or when X is known -for a stochastic volatility model
we never know the volatility with certainty.

The advantages of local volatility models are that firstly, no
additional (or untradable) source of randomness is introduced into
the model.  Hence the models are complete, unlike stochastic
volatility models. It is theoretically possible to perfectly hedge
contingent claims. Secondly, local volatility models can be also
calibrated to perfectly fit empirically observed implied
volatility surfaces, enabling consistent pricing of the
derivatives (an example is given in \cite{dupire1997pah}).
Thirdly, the local volatility model is able to account for a
greater degree of empirical observations and theoretical arguments
on volatility than time dependent volatility (for instance the
leverage effect). We will now look at some common local volatility
models.

\subsubsection{Constant Elasticity of Variance Model (CEV)}
%Price dependent models are also known as "level" dependent models
CEV was proposed by Cox and Ross \cite{cox1976ValCEV}:
%\cite{HULLIntro,p456}, BETTER MUSIELI P237:\cite{musiela2005mmf} VARIETY CEV REPRESENTATIONS EXIST
\begin{eqnarray}
\dfrac{dX}{X} & = & \mu dt+ \sigma(X)dW,\\
%\sigma(X) & = &  \alpha X^{n},\qquad \mbox{ for } n \in
%\mathbb{R^{+}}, \alpha \geq 0
\sigma(X) & = &  aX^{n-1},\label{CEV eqn2}
%\qquad \mbox{ for }0 \leq
%\beta \leq 1, \alpha > 0
\end{eqnarray}
%When n=1 CEV reduces to Geometric Brownian motion, when $n<1$
%($n>1$) the volatility increases (decreases) as stock price
%decreases, capturing the leverage effect.
where $0 \leq n \leq 1, a > 0$ are constants. For $n=0$ we obtain
Bachelier's model (equation (\ref{Bachelier eqn})) while $n=1$
gives the Geometric Brownian Motion model (equation (\ref{GBM
eqn})). Therefore n can be seen as a parameter for intuitively
choosing between the two extreme models. Additionally, n captures
the level of leverage effect since volatility increases as stock
price decreases, as can be seen from equation (\ref{CEV eqn2}).

The CEV model is analytically tractable, %it has a closed form solution
% for European options \cite{HULLIntro}.%HULL p457
 which is in contrast to the local volatility model in section
\ref{Implied Local Volatility section} where numerical solutions
for derivatives become analytically intractable
\cite{wilmott1998dta}. Additionally, by  %p290
appropriate choices of a and n  we can fit CEV to volatility
smiles (see for instance Beckers \cite{beckers1980cev}).

The CEV model has been developed over the years giving various CEV
modified models. For instance the square root CEV model
\cite{cox1976ValCEV}, which is similar to the CIR interest rate
model \cite{cox1985tts}, Schroder \cite{schroder1989cce}
re-expresses the CEV model in terms of a chi-squared distribution,
enabling derivation of  closed form solutions. %and an efficient algorithm for computing the probability density function.
Hsu et al. \cite{hsu2007cev} determine the CEV model's probability
density function while Lo et al. \cite{lo2000cev} derive the
option pricing formula for CEV with time dependent parameters.

\subsubsection{Mixture Distribution Models}\label{Mixture Distribution
Models} Mixture distributions have been applied to numerous
statistical applications for many years outside financial
mathematics (Everitt \cite{everitt1981fmd} gives a survey of such
applications). It had been known for some time that one could
capture many empirical characteristics of stock prices such as
volatility smiles through mixture distributions (see for instance
Bingham and Kiesel \cite{bingham2002spm} and Melick and Thomas
\cite{melick1997ras}). However Brigo and Mercurio
\cite{brigo2000mus} were the first to prove the theoretical
relation between volatility and a mixture of lognormal
distributions.

%Alexander presentation "volatility and risk"
Let us assume the risk neutral probability density function
$p(t,X)$ at time t of a stock price X(t) is a weighted sum of N
lognormal probability densities $\widetilde{p}_{i}(t,X)$:
\begin{eqnarray}
p(t,X) &=& \sum^{N}_{i=1}\tilde{w}_{i}(t)\widetilde{p}_{i}(t,X),\\
\mbox{where }\sum^{N}_{i=1}\tilde{w}_{i}(t) &=& 1,
\tilde{w}_{i}(t)\geq 0, \forall t.
\end{eqnarray}
The term $\tilde{w}_{i}(t)$ is a weighting for each component i at
a point in time t, $\widetilde{p}_{i}(t,X)$ denotes the
probability density of component i at a specific point in time for
a specific stock price X(t). We assume each
$\widetilde{p}_{i}(t,X)$ has the same mean $\mu$ but different
variances $\sigma_{i}^{2}(t)$.  Brigo and Mercurio then proved
that the stock price follows the process:
\begin{eqnarray}
dX/X &=& \mu dt+\sigma(X,t)dW,\\
\mbox{where }\sigma^{2}(X,t) &=& \sum^{N}_{i=1}{w}_{i}(t)
\sigma^{2}_{i}(t).
\end{eqnarray}
%NOT SURE SO LEAVE BELOW:
% Additionally the price of options is simply a weighted sum of
%Black-Scholes options:
%$$C(t,X,K,T,\sigma)=\sum^{M}_{i=1}w_{i}C_{BS}(t,X,K,T,\sigma_{i})$$

The mixture distribution model has been developed further by Brigo
et al.  (see for instance \cite{brigo2002lmd},\cite{brigo2003aap},
\cite{brigo2004amm}). Alexander has also developed the mixture
distribution model;  Alexander has applied it to the areas of
stress testing portfolios \cite{alexander2008dst}, combining it
with GARCH processes \cite{alexander2006nmg} and bivariate option
pricing \cite{alexander2004bnm}. GARCH processes will be covered
in more detail in section \ref{Other Volatility Models section}.

The mixture distribution models have been  used to model more
complicated volatility models due to their ability to capture a
variety of distributions. For example, Leisen \cite{leisen2005mld}
shows how a mixture distribution model approximates Merton's jump
diffusion model (to be covered in section \ref{Other Volatility
Models section}) amongst others, Lewis \cite{lewis2002mas} shows
how mixture models can approximate stochastic volatility models.

\subsubsection{Implied Local Volatility}\label{Implied Local Volatility section}
Dupire \cite{DUPILoc}, Derman and Kani \cite{derman1994rs} proved
for local volatility that a \textit{unique} risk-neutral
process existed that was consistent with option data. %, which we will now outline. %GATHERAL, LECTURE 1
The  price of a European call option by risk neutral valuation is
given by:
\begin{eqnarray}
C
 &=& e^{-rT}E^{\mathbb{Q}}[X(T)-K]^{+},\\
% &=& e^{-rT}\int_{0}^{\infty}max(X(T)-K,0)p(X(T))dX,\\
 &=& e^{-rT}\int_{K}^{\infty}(X(T)-K)p(X(T))dX, \label{BL part diff}
\end{eqnarray}
where p(X(T)) is the risk neutral probability density function for
X(T). Breeden and Litzenberger \cite{breeden1978psc} then showed
from equation (\ref{BL part diff}) that the risk neutral
cumulative distribution function $F(\cdot)$ at K is:
\begin{eqnarray}
\frac{\partial C}{\partial K} &=& -e^{-rT}F(X(T)\geq K).
\end{eqnarray}
Furthermore, the risk neutral probability density function
p(X(T)=K) is
\begin{eqnarray}\label{dupire BL eqn}
\frac{\partial^{2}C}{\partial K^{2}}e^{rT}=p(X(T)=K).
\end{eqnarray}
%%%%%%%%%%%%%%%%%%%%%%%%%%%%%%%%%%%%%%%%%%%%%%%%%%%%%%%%%%%%%%%%%%%%%%%%%%%%%%%%%%%%%%%%%%%%%%%%%%%%%%%%%%%%%
\begin{comment}
Breeden and Litzenberger \cite{breeden1978psc} then showed by
partially differentiating equation (\ref{BL part diff}) with
respect to K gives the risk neutral cumulative distribution
function $F(\cdot)$ at K:
\begin{eqnarray}
C
&=& e^{-rT}\int_{K}^{\infty}(X(T)-K)p(X(T))dX,\\
  &=&
  e^{-rT}\int_{K}^{\infty}X(T)p(X(T))dX-e^{-rT}\int_{K}^{\infty}Kp(X(T))dX,\\
 \frac{\partial C}{\partial K}  &=&
-e^{-rT}\int_{K}^{\infty}p(X(T))dX,\\
 &=& -e^{-rT}F(X(T)\geq K).
\end{eqnarray}
%definition of CDF http://en.wikipedia.org/wiki/Cumulative_distribution_function
Furthermore, partially differentiating equation (\ref{BL part
diff}) twice with respect to K gives the risk neutral probability
density function p(X(T)=K):
\begin{eqnarray}\label{dupire BL eqn}
\frac{\partial^{2}C}{\partial K^{2}}e^{rT}=p(X(T)=K).
\end{eqnarray}

\end{comment}
%%%%%%%%%%%%%%%%%%%%%%%%%%%%%%%%%%%%%%%%%%%%%%%%%%%%%%%%%%%%%%%%%%%%%%%%%%%%%%%%%%%%%%%%%%%%%%%%%%%%%%%%%%%%%
Hence we can recover the risk neutral density p(X(T)) from option
data. This probability can be interpreted as the current view of
the future outcome of the stock price. % %\cite{wilmott1998dta, p292}

%\cite{wilmott1998dta, p294}
Dupire \cite{DUPILoc} then showed by applying p(X(T)) from
equation (\ref{dupire BL eqn})  (obtained from Breeden and
Litzenberger's work) to the Fokker-Planck equation, %(also known as the forward Kolmogorov equation)
%with the initial condition C(K,0)=max(S(0)-K,0)
one could obtain Dupire's equation:
%although \sigma(K,T)=\sigma(X,T) according to wilmott p294, best leave since not understand
%MADE SUBSTITUTION K=X, MAKE SENSE FROM BREEDEN LITZ EQN.BOTH ARE EQUAL.
\begin{eqnarray}\label{Dupire eqn1}
\dfrac{\partial C}{\partial T} %& =& \sigma^2.\dfrac{K^{2}}{2}.\dfrac{\partial^{2} C}{\partial
%K^{2}}-(r-D)K.\dfrac{\partial C}{\partial K}-DC,\\
 & =& \sigma^2(X,T).\dfrac{X^{2}}{2}.\dfrac{\partial^{2}
C}{\partial X^{2}}-(r-D)X.\dfrac{\partial C}{\partial X}-DC,
\end{eqnarray}
where D is the dividend. Rearranging equation (\ref{Dupire eqn1})
gives:
%\begin{eqnarray}\label{Dupire eqn2}
%\sigma=\sqrt{\dfrac{\dfrac{\partial C}{\partial
%T}+(r-D)K\dfrac{\partial C}{\partial
%K}+DC}{\dfrac{K^2}{2}\dfrac{\partial^2 C}{\partial K^2}}}.
%\end{eqnarray}
\begin{eqnarray}\label{Dupire eqn2}
\sigma(X,T)=\sqrt{\dfrac{\dfrac{\partial C}{\partial
T}+(r-D)X\dfrac{\partial C}{\partial
X}+DC}{\dfrac{X^2}{2}\dfrac{\partial^2 C}{\partial X^2}}}.
\end{eqnarray}
Therefore the local volatility $\sigma(X,T)$ can be fully
extracted from option data.

It can be seen from equation (\ref{Dupire eqn2}) that calculating
$\sigma$ requires partial differentials with respect to T and K.
We therefore require a continuous set of options data for all K
and T. This is highly unrealistic and quoted option prices tend to
suffer from significant illiquidity effects, affecting option
bid-ask spreads \cite{norden2003aop}. Furthermore, Pinder
\cite{pinder2003eei} shows that option bid-ask spreads are related
to volatility, expiry and trading volume. Since option data is
discrete we require some interpolation method to convert it to
continuous data, for example Monteiro et al.
\cite{monteiro2008rrn} apply a cubic spline method. However,
Wilmott
\cite{WILLQuantFin1} %,25.6
states that local volatility computation is highly sensitive to
interpolation methods.
%DON'T UNDERSTAND:
% Also, it can be seen from equation
%\ref{Dupire eqn2} for $K>>S$ or $S>>K$ lead to extremely small
%numerator and denominator figures, causing computational problems.
%The reader is referred to Wilmott \cite{WILLQuantFin1} for further
%derivation.

Dupire's model implicitly assumes the options data contains all
the information on the underlying's volatility if we calibrate to
options data alone. However there is evidence to show historic and
implied volatility differ significantly \cite{christensen1998rbi}.
Furthermore, calibrating the local volatility surface to the
options data tends to be unstable with time,  since the surface
significantly changes from one week to another
\cite{dumas1998ivf}. %\cite{fouque2000dfm,p39}

Numerical computation of local volatility has been implemented by
Andersen and Brotherton-Ratcliffe using a finite difference method
\cite{andersen1998eov}. Derman and Kani (\cite{derman1996itt},
\cite{derman1996lvs}) and Rubinstein \cite{rubinstein1994ibt}
determine local volatilities by fitting a unique binomial tree to
the observed option prices. Tree fitting also has the
computational advantage of not being affected by different
interpolation methods.

\subsection{Stochastic Volatility Models}\label{stochastic volatility
models}
\subsubsection{Definition and Characteristics}\label{Stochastic Volatility Model Definition and Characteristics section}
Although local volatility models were an improvement on time
dependent volatility, they possessed certain undesirable
properties. For
 example, volatility is perfectly correlated (positively or negatively) with
stock price %\cite{fouque2000dfm,p39}
yet empirical observations suggest no perfect correlation exists. % \cite{hobson1996sv}
 Stock prices empirically exhibit volatility clustering but under
local volatility this does not necessarily occur. Consequently
after local volatility development, models were proposed that
allowed volatility to be governed by its own stochastic process.
We now define stochastic volatility.
%\cite{fouque2000dfm,p40}:
\begin{defi}
Assume X follows the stochastic differential equation
\begin{eqnarray}
 dX/X &=& \mu dt + \sigma(\omega)dW_{1}.
%d\sigma_{1}/\sigma_{1} &=& \mu_{2}dt+ \sigma_{2}dW_{2}
\end{eqnarray}
Volatility is stochastic if $\sigma(\omega)$ is governed by a
stochastic process that is driven by another (but possibly correlated) random process, typically another Wiener process $dW_{2}$. % FOUQUE BOOK P41:
The probability space $(\Omega,\mathcal{F},\mathbb{P})$ is
$\Omega=\mathcal{C}([0,\infty):\mathbb{R}^{2})$, with filtration
$\{\mathcal{F}_{t}\}_{t\geq 0}$ representing information on two
Wiener processes $(W_{1},W_{2})$.\\
The process governing $\sigma(\omega)$ must always be positive for
all values since volatility can only be positive.
%%%%%%%%%%%%%%%%%%%%%%%%%%%%%%%%%%%%%%%%%%%%%%%%%%%%%%%%%%%%%%%%%%%%%%%%%%%%%%%%%%%%%%%%%%%%%%%%%%%%%%%%%%%%%%%%%%%%%%%%%%%%
\begin{comment}
%\cite{fouque2000dfm,p40}
\begin{itemize}
%  \item satisfy regularity conditions -LIPSHITZ, GROWTH,
 % OKSENDAL
  \item must always be positive for all values since volatility can only be
  positive;
  \item does not need to be an Ito process (e.g. jump processes and
  Markov chains are acceptable).
\end{itemize}
\end{comment}
%%%%%%%%%%%%%%%%%%%%%%%%%%%%%%%%%%%%%%%%%%%%%%%%%%%%%%%%%%%%%%%%%%%%%%%%%%%%%%%%%%%%%%%%%%%%%%%%%%%%%%%%%%%%%%%%%%%%%%%%%%%%%
The Wiener processes have instantaneous correlation $\rho$ $\in
[-1,1]$ defined by:
\[
corr(dW_{1}(t),dW_{2}(t))=\rho dt.
\]
\end{defi}% FOUQUE BOOK P41:
Empirically $\rho$ tends to be negative in equity markets due to
the leverage effect (see section \ref{Empirical Characteristics of
Volatility}) but close to 0 in
the currency markets. %\cite{FOUQDer}.
Although $\rho$ can be a function of time we assume it is a
constant throughout this thesis.
%Note that we can re-express:
%\begin{eqnarray}
%dW_{2} = \rho W_{1}+\sqrt{1-\rho^2}W_{2}
%\end{eqnarray}

The key difference between local and stochastic volatility is that
local volatility  is not driven by a random process of its own;
there exists only one source of randomness ($dW_{1}$). In
stochastic volatility models, volatility has its own source of
randomness ($dW_{2}$) making volatility intrinsically stochastic.
We can therefore never definitely determine the volatility's
value, unlike in local volatility.
%Hence due to the introduction of another source of randomness,
%stochastic volatility models are 2 factor models.
%Thus the price SDE becomes:
%\begin{eqnarray*}
%dS(t)&=& \mu S(t)dt+\sigma(\omega,t) S(t)dB_{1}(t)\\
%\end{eqnarray*}

The key advantages of stochastic volatility models are that they
capture a richer set of empirical characteristics compared to
other volatility models \cite{musiela2005mmf}. Firstly, stochastic
volatility models generate return distributions similar to what is
empirically observed.
%\cite{fouque2000dfm},42.
For example, the return distribution has a fatter left tail and
peakedness
 compared to normal distributions, with tail asymmetry controlled by
 $\rho$ \cite{durham2007smm}. Secondly, Renault and Touzi \cite{renault1996oha} proved
 volatility that is stochastic and $\rho$=0 always produces implied volatilities that smile (note
that volatility smiles do not necessarily imply volatility is
stochastic).

Thirdly, historic volatility shows significantly higher
variability than would be expected from local or time dependent
volatility, which could be better explained by a stochastic
process. A particular case in point is the dramatic change in
volatility during the 1987 October crash (Schwert
\cite{schwert1990sva} gives an empirical study on this). Finally,
stochastic volatility accounts for the volatility's empirical
dependence on the time scale measured (as discussed in section
\ref{Empirical Characteristics of Volatility}), which should not
occur under local or time dependent volatility.
%In particular if volatility is a stochastic mean reverting
%process, we would expect volatility to become more Gaussian as the
%time scale is increased (Gaussian aggregation);  this is
%empirically exhibited by stock prices (see section \ref{Empirical
%Characteristics of Volatility}) and will be proven in section
%\ref{Stochastic Volatility and the Driving Process}.
%A brief proof is given in \cite{fouque2000dfm,p52-3}.
%For $\rho \neq$0 no analytical conclusion can be drawn about the
%implied volatility relation. The reader is referred to
%\cite{fouque2000dfm} chapter 2 %p53-5
%for more information.
%It is interesting to note that modelling volatility as stochastic
%is not noticeable by visual inspection of the stock price process
%-SEE $http://www.xplore-stat.de/tutorials/stfhtmlnode46.html$

The disadvantages of stochastic volatility are firstly that these
models introduce a non-tradable source of randomness, hence the
market is no longer complete and we can no longer uniquely price
options or perfectly hedge. Therefore the practical applications
of stochastic volatility  are limited. Secondly stochastic
volatility models tend to be analytically less tractable. In fact,
it is common for stochastic volatility models to have no closed
form solutions for option prices. Consequently option prices can
only be calculated by simulation (for example Scott's model in
section \ref{Significant Stochastic Volatility Models}).

\subsubsection{Stochastic Volatility and the Driving Process}\label{Stochastic Volatility and the Driving
Process} Stochastic volatility models  fundamentally differ
according to their driving mechanisms for their volatility
process. Different driving mechanisms maybe favoured due to their
tractability, theoretical or empirical appeal and we can
categorise stochastic volatility models according to them.
%%%%%%%%%%%%%%%%%%%%%%%%%%%%%%%%%%%%%%%%%%%%%%%%%%%%%%%%%%%%%%%%%%%%%%%%%%%%%%%%%%%%%%%%%%%%%%%%%%%%%%%%%%%%%%%%%%
\begin{comment}
 Fouque (\cite{FOUQDer},p42) conveniently
lists the most popular driving processes:
\\\\
\begin{tabular}
[l]{|p{3cm}|p{3cm}|p{2cm}|p{4cm}|}\hline Model & Correlation
$\rho$ & $\sigma$=f(y) & y process \\\hline Hull-White & 0 &
f(y)=$\sqrt{y}$ & lognormal \\\hline Scott & 0 & f(y)=$e^{y}$ &
Mean-reverting OU
\\\hline Stein-Stein & 0 & f(y)=$\mid y \mid$ & Mean-reverting OU
\\\hline Ball-Roma & 0 & f(y)=$\sqrt{y}$ & CIR\\\hline Heston &
$\rho\neq0$  & f(y)=$\sqrt{y}$ & CIR
\\\hline
\end{tabular}
\\\\\\
\end{comment}
%%%%%%%%%%%%%%%%%%%%%%%%%%%%%%%%%%%%%%%%%%%%%%%%%%%%%%%%%%%%%%%%%%%%%%%%%%%%%%%%%%%%%%%%%%%%%%%%%%%%%%%%%%%%%%%%%%
Many stochastic volatility models favour a mean reverting driving
process. A mean reverting stochastic volatility process is of the
form \cite{fouque2000dfm}:%p40:
\begin{eqnarray}
 \sigma &=& f(Y),\\
 dY &=& \alpha(m-Y)dt+\beta dW_{2},\label{mean
 revert eqn definition}
 \end{eqnarray}
where:
\begin{itemize}
 \item $\beta \geq 0$ and $\beta$ is a constant;
  \item m is the long run mean value of $\sigma$;
  \item $\alpha$ is the rate of mean reversion.
\end{itemize}
Mean reversion is the tendency for a process to revert around its
long run mean value. We can economically account for the existence
of mean reversion through the cobweb theorem, which claims prices
mean revert due to lags in supply and demand \cite{lipsey1992fpe}.
The inclusion of mean reversion ($\alpha$) within volatility is
important, in particular, it controls the degree of
volatility clustering (burstiness) if all other parameters are unchanged. Volatility %p68,87
 clustering is an important empirical characteristic of
many economic or financial time series \cite{engle1982ach}, which
neither local nor time dependent volatility models necessarily
capture.
%\cite{HULLIntro}\cite{FOUQDer,p.59}
Additionally, a high $\frac{1}{\alpha}$ can be thought of as the
time required to decorrelate or ``forget"
its previous value.%p66,87

The equation (\ref{mean revert eqn definition}) is an
Ornstein-Uhlenbeck process in Y with known solution:  %\cite{fouque2000dfm,p41}:
\begin{eqnarray}
Y(t) = m+(Y(0)-m)e^{-\alpha t}+\beta \int^{t}_{0}e^{-\alpha
t}dW_{2},
%
%removed s in \alpha(t-s) since i think just initial time from http://planetmath.org/encyclopedia/OrnsteinUhlenbeckProcess.html
%
%\sigma(t) = m+(\sigma(0)-m)e^{-\alpha t}+\beta
%\int^{t}_{0}e^{-\alpha(t-s)}dW_{2},
\end{eqnarray}
 where Y(t) has the
distribution
\begin{eqnarray}\label{OU dist eqn}
Y(t) \sim \mathcal{N} \left(m+(Y(0)-m)e^{-\alpha
t},\frac{\beta^2}{2\alpha}(1-e^{-2\alpha t})\right).
\end{eqnarray}
Note that alternative processes to equation (\ref{mean revert eqn
definition}) could have been proposed to define volatility as a
mean reverting stochastic volatility model, for example the Feller
or Cox-Ingersoll-Ross (CIR) process
\cite{fouque2000mrs}: %p8
\begin{eqnarray}
dY &=&  \alpha(m-Y)dt+\beta\sqrt{Y}dW_{2}.
\end{eqnarray}
However with $\sigma=f(Y)$ in equation (\ref{mean revert eqn
definition}) we can represent a broad range of mean reverting
stochastic volatility models in terms of a function of Y.
%%%%%%%%%%%%%%%%%%%%%%%%%%%%%%%%%%%%%%%%%%%%%%%%%%%%%%%%%%%%%%%%%%%%%%%%%%%%%%%%%%%%%%%%%%%%%%%%%%%%%%%%%%%%%%%%%%%%%%%%%%%%%%
\begin{comment}DON'T INCLUDE AS DON'T UNDERSTAND

%The OU can be geometric OU e.g. Wiggins model or  squared OU e.g.
%Stein and Stein model or inverse-gamma model and is the continuous
%time limit of the GARCH(1,1) model.
%NEED ADDRESS OF ISSUE OF WHY DIFFERENT PROCESSES CHOSEN -WHAT
%EMPIRICAL AND THEORETICAL EVIDENCE SUPPORT AND WHAT PROCESS MEAN.
\begin{itemize}
  \item
\item Feller or Cox-Ingersoll-Ross (CIR) process:
\begin{eqnarray}
d\sigma &=&  \alpha(m-\sigma)dt+\beta\sqrt{\sigma}dW_{2}.
\end{eqnarray}
\item Squared volatility process:
\begin{eqnarray}
d\sigma^{2} = \alpha(m-\sigma^{2})dt+\sigma \beta dW_{2}.
\end{eqnarray}
%\item the lognormal process: -NOT MEAN REVERTING, FOUQUE P40
%\begin{eqnarray*}
%d\sigma/\sigma &=& adt+bdW_{2}
%\end{eqnarray*}

\end{itemize}
\end{comment}
%%%%%%%%%%%%%%%%%%%%%%%%%%%%%%%%%%%%%%%%%%%%%%%%%%%%%%%%%%%%%%%%%%%%%%%%%%%%%%%%%%%%%%%%%%%%%%%%%%%%%%%%%%%%%%%%%%%%%%%%%%%%%%

\subsubsection{Significant Stochastic Volatility Models} \label{Significant Stochastic Volatility
Models} There is no generally accepted canonical stochastic
volatility
model %\cite{fouque2000dfm,p56}
and a large number of them exist, therefore we review here the
most significant ones.

\begin{enumerate}
\item Johnson and Shanno Model\\
Johnson and Shanno in 1987 \cite{johnson1987opv}  proposed one of
the first stochastic volatility models:
\begin{eqnarray}
 dX &=& \mu_{1} Xdt+\sigma X^{n}dW_{1},\qquad n \geq 0,\\
   d\sigma &=& \mu_{2}\sigma dt+\sigma^{k}\beta dW_{2}, \qquad k \geq
   0,
% dX &=& \phi Xdt+\sigma X^{n}(t)dW_{1},\qquad n \geq 0,\\
%   d\sigma &=& \mu\sigma dt+\xi\sigma^{m}dW_{2}, \qquad m \geq 0.
\end{eqnarray}
where $corr(dW_{1}(t),dW_{2}(t))=\rho dt$. A Monte Carlo method is
proposed to determine the price of options under the stochastic
volatility process by risk neutral valuation. Johnson's and
Shanno's computational results show that their option prices are
consistent with what is empirically observed (see section
\ref{Empirical Characteristics of Volatility}), that is they
exhibit a volatility smile and an increase in value with expiry.

%%%%%%%%%%%%%%%%%%%%%%%%%%%%%%%%%%%%%%%%%%%%%%%%%%%%%%%%%%%%%%%%%%%%%%%%%%%%%%%%%%%%%%%%%%%%%%%%%%%%%%%%%%%%
\begin{comment}
 By applying the Ito's lemma one can the call option SDE. To eliminate the 2 random
terms one of the following assumptions are made:
\begin{enumerate}
\item risk due stochastic variance is completely diversifiable by
buying a stock and shorting a call option on that stock. They do
not explain how this is achieved
\item
\end{enumerate}
\end{comment}
%%%%%%%%%%%%%%%%%%%%%%%%%%%%%%%%%%%%%%%%%%%%%%%%%%%%%%%%%%%%%%%%%%%%%%%%%%%%%%%%%%%%%%%%%%%%%%%%%%%%%%%%%%%%
\item Scott Model\\
Scott in 1987 \cite{scott1987opv} considered the case where one
assumes a geometric process for stock prices and an
Ornstein-Uhlenbeck process for the volatility:
\begin{eqnarray}
 dX/X &=& \mu dt+\sigma dW_{1},\\ %PROCESS TAKEN FROM ORIGINAL PAPER. ORIGINAL PAPER LOOKS AT EXPOU AND OU
   d\sigma &=& \alpha(m-\sigma)dt+\beta dW_{2}.
\end{eqnarray}
Scott proposed a stochastic mean reverting process based on
empirical stock price returns and describes a parameter estimation
method based on moments matching. Scott assumes
%a zero market price of volatility risk and
$\rho$=0 %p7 of paper
to facilitate computation of option prices. The option prices for
Scott's process are calculated by Monte Carlo simulation and he
observes a marginal improvement in option pricing accuracy
compared to standard Black-Scholes option pricing.
%%%%%%%%%%%%%%%%%%%%%%%%%%%%%%%%%%%%%%%%%%%%%%%%%%%%%%%%%%%%%%%%%%%%%%%%%%%%%%%%%%%%%%%%%%%%%%%%%%%%%%%%%%%%%%%%%
\begin{comment}
\item Wiggins' Model\\
 Wiggins in 1987 \cite{wiggins1987ovu} proposed another
volatility models:
\begin{eqnarray*}
 dX/X &=& \mu dt+\sigma dW_{1}\\
   d\sigma &=& f(\sigma)dt+\xi\sigma dW_{2}
\end{eqnarray*}
Note that $dW_{1}(t),dW_{2}$ are correlated by $\rho$ and
$f(\sigma)$ can follow e.g. mean reverting process. By an
arbitrage argument, Wiggins derives the PDEs satisfied by call
option prices.
\end{comment}
%%%%%%%%%%%%%%%%%%%%%%%%%%%%%%%%%%%%%%%%%%%%%%%%%%%%%%%%%%%%%%%%%%%%%%%%%%%%%%%%%%%%%%%%%%%%%%%%%%%%%%%%%%%%%%%%%
\item Hull-White Model\\
Hull and White modelled volatility as follows \cite{hull1987poa}:
\begin{eqnarray}
dX/X=\mu_{1} dt+ \sigma dW_{1},\\
d\sigma^{2}/\sigma^{2}=\mu_{2}dt +\beta dW_{2}.
\end{eqnarray}
The Hull-White model is an important contribution since it
provides a closed form solution to European option prices when
$\rho$=0  and for a given risk neutral measure $\gamma$. Let us
define
\begin{eqnarray}
\widetilde{\sigma^{2}}=\frac{1}{T-t}\int_{t}^{T} \sigma^{2}(s) ds,
\end{eqnarray}
where $\widetilde{\sigma^{2}}$ is a random variable. The option
price is computed using the standard Black-Scholes formula, under
a risk neutral measure $\gamma$, with volatility
$\sqrt{\widetilde{\sigma^{2}}}$. In other words:
\begin{eqnarray}\label{HW option eqn}
C(t,X,K,T,\sigma(\omega))=E^{\gamma}[C_{BS}(t,X,K,T,\sqrt{\widetilde{\sigma^{2}}})].
\end{eqnarray}
The option pricing equation (\ref{HW option eqn}) provides results
consistent with empirically observed currency options, where
empirically
$\rho\simeq 0$.  % \cite{HULLIntro}.
Furthermore equation (\ref{HW option eqn})  is still valid for any
stochastic volatility process provided $\rho=0$
\cite{fouque2000dfm}. For correlated volatility, option prices are
obtained using Monte
Carlo simulation. %but in the case $\alpha\sim 0$
\item Stein and Stein  Model\\
 Stein and Stein in 1991 \cite{stein1991spd} proposed an
Ornstein-Uhlenbeck process for volatility based on tractability
and empirical considerations:
\begin{eqnarray}
 dX/X &=& \mu dt+\sigma dW_{1},\\
   d\sigma &=& -\alpha(\sigma-m)dt+\beta dW_{2},\label{SteinStein vol eqn}
\end{eqnarray}
where $\rho=0$. Note that although equation (\ref{SteinStein vol
eqn}) implies volatility can be negative, Stein and Stein state
that only $\sigma^{2}$ is ever applied in calculation \cite{stein1991spd}. A %p729.
 closed form solution to option prices is derived for particular
 choices of risk neutral measures, which is in contrast to Johnson and Shanno \cite{johnson1987opv} and Scott \cite{scott1987opv}, %Wiggins
who only provide numerical methods to option pricing.
%SEE WILMOTT FOR MORE INFO*******\\
%http://www.quantlet.com/mdstat/scripts/stf/html/stfhtmlnode45.html\newline
%http://www.xplore-stat.de/tutorials/stfhtmlnode46.html#hes_model
\item Heston model\\
Heston's model \cite{heston1993cfs} created in 1993 stands out
from other stochastic volatility models because there exists an
analytical solution for European options that takes in account
correlation between  $dW_{1}$ and $dW_{2}$ (although this requires
assumptions on the risk neutral measure):
%It was also one of the first models that was able to
%explain the
%smile and simultaneously allow a front-office implementation and a
%market consistent valuation of many exotics. The Heston model can
%be applied to pricing options on bonds and FX if interest rates
%are assumed to be stochastic, however parameter estimation is
%significant.
\begin{eqnarray}
 dX/X &=& \mu dt+\sigma dW_{1},\\
   d\sigma^{2} &=& \alpha(m-\sigma^{2})dt+\beta \sigma dW_{2}.%\\
   %&=& (\alpha m-\alpha \sigma^{2})dt+\beta \sigma dW_{2}.
\end{eqnarray}
%We call  $\beta$ the volatility of volatility and
The $d\sigma^{2}$ modelling process originated from CIR interest
rate model \cite{cox1985tts}. To price an option under risk
neutral valuation, we must specify a risk neutral measure (due to
market incompleteness) and this is chosen on economic
justifications \cite{heston1993cfs}.
%HestonExplained.pdf
%To price an option under risk neutral valuation, we require a risk
%neutral measure which specifies the market price of volatility
%risk $\gamma$ (due to incompleteness).This is assumed to be
%linearly related to the instantaneous variance due to economic
%reasons \cite{heston1993cfs}, that is $\gamma =k\sigma^{2}$ where
%k is an arbitrary constant.
%%%%%%%%%%%%%%%%%%%%%%%%%%%%%%%%%%%%%%%%%%%%%%%%%%%%%%%%%%%%%%%%%%%%%%%%%%%%%%%%%%%%%%%%%%%%%%%%%%%%%%%%%%%
\begin{comment}DON'T UNDERSTAND SO LEAVE
Following Musiela and Rutkowski
\cite{musiela2005mmf} %p263
the risk neutral process is now
\begin{eqnarray*}
 dX/X &=& r dt+\sigma dW_{1}^{\mathbb{Q}},\\
   d\sigma^{2} &=& \alpha^{*}(m^{*}-\sigma^{2})dt+
   \beta \sigma dW_{2}^{\mathbb{Q}},\\
  corr(dW_{1}^{\mathbb{Q}}(t),dW_{2}^{\mathbb{Q}}(t)) &=& \rho dt,
\end{eqnarray*}
where
\begin{itemize}
  \item $\alpha^{*}=\alpha +\gamma=\alpha +k\sigma^{2}$;
  \item $m^{*}=\dfrac{\alpha m}{\alpha +k\sigma^{2}} =\dfrac{\alpha
m}{\alpha +\gamma}$.
\end{itemize}

\end{comment}
%%%%%%%%%%%%%%%%%%%%%%%%%%%%%%%%%%%%%%%%%%%%%%%%%%%%%%%%%%%%%%%%%%%%%%%%%%%%%%%%%%%%%%%%%%%%%%%%%%%%%%%%%%%
Heston then finds an analytical solution for options using Fourier
transforms;  the reader is referred to Heston \cite{heston1993cfs}
and Musiela and Rutkowski
\cite{musiela2005mmf} for a derivation. %p263+

Due to the existence of an analytic solution to option pricing the Heston model has been subsequently % Motivated by Breeden's consumption based model 1979 and CIR 1985 intermporal equilibrium model .
developed by various researchers. For example, Scott includes
stochastic interest rates \cite{scott1997pso}, Pan includes
stochastic dividends \cite{pan2002jrp} and Bates adds jumps to the
stochastic process \cite{bates1996jas}. %NOT TO SV PROCESS

\end{enumerate}
%******SEE GATHERAL P13+ FOR DERIVATION ON OPTION PRICE DERIVATION????

\subsection{Other Volatility Models}\label{Other Volatility Models
section}
%\subsection{Merton's Jump Diffusion Model}
Apart from Bates \cite{bates1996jas}, the models discussed so far
all have continuous sample paths yet empirical stock prices appear
to exhibit ``jumps". Additionally, the frequency and magnitude of
the jumps are too large to be explained by lognormal
distributions, which would be obtained under
a GBM process. %\cite{wilmott1998dta,p253}
Merton therefore proposed the jump diffusion model
\cite{merton1975opu}, which added a random jump
component to the GBM process:  %\cite{HULLIntro,p457}:
%INTRODUCES INCOMPLETENESS TOO, GIRSANOV THEOREM FOR JUMPS TOO
\begin{eqnarray}
dX/X &=& (\mu -\theta k)dt+\sigma dW+d\mathcal{P},
\end{eqnarray}
where
\begin{itemize}
  \item $\theta$ is the average number of jumps/year;
  \item k is the average jump size measured as a percentage of the asset price X;
  \item $\mathcal{P}$ is a Poisson process and independent of dW. %Poi($\lambda$)
\end{itemize}
A Poisson process counts the number of events that occur in a
given period. We define a Poisson process
$\mathcal{P}(\mathcal{\nu})$ as  \cite{lindgren1966ipa}: %p68
%\cite{kreyszig:ims},http://en.wikipedia.org/wiki/Poisson_process
\begin{eqnarray}
p(X=k)=\dfrac{e^{-\mathcal{\nu}t}(\mathcal{\nu}t)^{k}}{k!},k=0,1,2,....
\end{eqnarray}
where
\begin{itemize}
\item $\mathcal{\nu}$ is called the rate parameter, where $\mathcal{\nu}$ is
the expected number of events per unit time;
\item X is a random variable denoting the number of events;
\item p(X=k) is the probability that the random number of events from 0 to time
t equals k.
\end{itemize}
In the case of the jump diffusion model the events are the jumps
themselves.
%The probability of a jump in time dt is $\theta dt$ with the
%average growth rate arising from the jumps alone is $\theta k$.
To obtain a closed form solution for option prices under the jump
diffusion model, Merton makes a key argument as follows.
%HULLp458
We know that a replicating portfolio constructed in the derivation
of the original Black-Scholes equation (\cite{black1973poa}) will
eliminate the risk arising from the GBM and so must earn the risk
free rate of return. Next Merton assumes the jump process
represents nonsystematic risk and risk that is not \textit{priced
into} the market. Therefore the same replicating portfolio applied
in the Black-Scholes equation must earn the risk free rate of
return.  Merton then proved that the option price is the sum of an
infinite series of options, valued by the standard Black-Scholes
option pricing equation without the jumps.
\begin{eqnarray*}
C(X(t),K,t,T,r,\sigma) & =&
\sum_{n=0}^{\infty}\dfrac{e^{-\theta'\tau}(\theta'\tau)^{n}}{n!}C_{BS}(X(t),K,t,T,r_{n},\sigma_{n}),\\
\mbox{where}\\
\tau &=& T-t,\\
\theta'&=& \theta(1+k),\\
r_{n} &=& r-\theta k+\dfrac{nln(1+k)}{\tau},\\
\sigma_{n}^{2} &=& \sigma^{2}+\dfrac{n\varsigma^{2}}{\tau}.
\end{eqnarray*}

The $n^{th}$ option in the series is valued by the standard
Black-Scholes equation, $C_{BS}$, assuming n jumps have occurred
before expiry. %The reader is referred
%to\cite{wilmott1998dta,p330} for the full equation.
Since the series converges exponentially it can be implemented
computationally \cite{cont2004fmj}. The term $\varsigma^{2}$
arises from the fact that a Poisson process can be approximated by
a normal distribution. Therefore we can assume the logarithm of
the jumps are normally distributed with variance $\varsigma^{2}$.
The jumps fatten the return distribution's tail, therefore the
model is more consistent with empirically observed distributions
compared to GBM.
% NOT SURE: and account for the volatility smiles and according to FIRST ASIA PAPER fits extremely well for most equity derivatives.
Merton accounts for jumps as the arrival of new information that
have more than a marginal effect on price movements.

%\textbf{Implied Volatility Models}\\
%http://www.quantlet.com/mdstat/scripts/stf/html/stfhtmlnode45.html\newline
%REF: WILMOTT QAUNT FIN CHAPTER 26\\
Another class of volatility models is to use a stochastic process
to model the evolution of the implied volatility directly. In all
the models discussed so far, we have assumed implied volatility
and the underyling's volatility are the same. Implied volatility
methods model an option's (or any derivative's) implied volatility
$\sigma^{\ast}$ separately from the underlying's volatility
$\sigma$. One significant implied volatility model is
Schonbucher's stochastic implied volatility model
\cite{schonbucher1999mms}. Schonbucher models the implied
volatility $\sigma^{\ast}$ of a vanilla option as a SDE and the
underlying's SDE separately:
\begin{eqnarray}
d\sigma^{\ast}=\mu_{2}dt+\sigma_{2}dW_{1}+\beta dW_{2}.
\end{eqnarray}
The underlying's SDE is:
\begin{eqnarray}
dX/X=\mu_{1} dt+\sigma_{1}dW_{1}.
\end{eqnarray}

Another class of volatility models is the lattice approach, with
its origins from binomial trees by Cox et al. \cite{cox1979ops}.
Britten-Jones and Neuberger \cite{brittenjones2000opi} model
stochastic volatility by fitting a lattice model that is
consistent with observed option prices. The model is parameterised
by up and down price movements and their associated probabilities
for each branch, using empirical option data as its input. This is
an improvement on Derman's and Kani's model, who fit a tree to
local volatility only. Britten-Jones and Neuberger also show how a
variety of stochastic volatility models (such as regime switching
volatility) can be fitted to be consistent with observed option
price data.

Discrete time volatility models are another class of models that
exist, such as
 GARCH(p,q) \cite{bollerslev1986gac}, the generalised autoregressive
  conditional heteroscedasticity model:  %and ARV
\begin{eqnarray}
{\sigma_{t}}^{2}=a_{0}+ {\displaystyle\sum\limits_{i=1}^{q}}
a_{i}{\varepsilon}_{t-{i}}^{{2}}+
{\displaystyle\sum\limits_{i=1}^{p}}
b_{i}{\sigma}_{t-{i}}^{{2}},\mbox{ for }\varepsilon_{t}\sim
\mathcal{N}(0,\sigma_{t}^{2}),
\end{eqnarray}
where $a_{i}$ and $b_{i}$ are weighting constants. We obtain the
ARCH(q) \cite{engle1982ach} model by setting p=0. Due to the
success of GARCH in econometrics
it has been substantially extended by various %http://en.wikipedia.org/wiki/GARCH
researchers, for example Nelson \cite{nelson1991cha}, Sentana
\cite{sentana1995qam} and Zakoian \cite{zakoian1994thm}. Whereas
with continuous time modelling one is able to derive closed form
solutions, which can reduce computation and provide new insights,
this is generally not possible with discrete time models.
% We can also apply mathematical theories such as stochastic calculus to
%derive important results.
%NOT CONVINCED BELOW
%Furthermore continuous time models deal with time intervals in a
%consistent manner whereas discrete models are unable to do this
%because price measurements tend to be uneven in time e.g. stock
%exchanges close during bank holidays, weekends etc... , prices may
%be taken at the middle of the day or end etc... .

Finally, one class of volatility models are ``hybrid" models;
combining various volatility models into one. For instance
Alexander \cite{alexander2004nmd} extends Brigo's and Mercurio's
mixture model \cite{brigo2000mus} by combining it with the
binomial tree of Cox et al. \cite{cox1979ops} to incorporate
stochastic volatility. The SABR model \cite{hagan2002msr} is a
stochastic extension of the CEV model \cite{cox1976ValCEV}, where
SABR is an abbreviation for stochastic alpha, beta and rho in its
equations. The SABR model captures the dynamics of a forward price
F of some asset (e.g. stock) under stochastic volatility; its risk
neutral dynamics under measure $\mathbb{Q}$ is:
%http://en.wikipedia.org/wiki/SABR_Volatility_Model
\begin{eqnarray}
dF(t) &=& \sigma(t) F^{\beta}(t)dW_{1}^{\mathbb{Q}},\\
d\sigma(t) &=&
\alpha\sigma(t)dW_{2}^{\mathbb{Q}}, \\
\mbox{where }corr(dW_{1}.dW_{2}) &=& \rho \in [-1,1], 0 \leq \beta
\leq 1, \alpha \geq 0 .
\end{eqnarray}

A common shortfall in all the volatility models reviewed so far
has been that all these models are short term volatility models.
The models implicitly ignore any long term or broader economic
factors influencing the volatility model, which is empirically
unrealistic and theoretically inconsistent. Furthermore, although
some models may specify a closed form solution for option pricing,
they provide no method or recommendation for calibration, which is
important to modelling and option pricing.

\section{Conclusions}
This paper has surveyed the key volatility models and
developments, highlighting the innovation associated with each new
class of volatility models. In conclusion it can be seen from our
 review of volatility models that the development of
has progressed in a logical order to address key shortcomings of
previous models.

Time dependent models addressed option prices varying with
expiration dates, local volatility also addressed volatility
smiles and the leverage effect, whereas stochastic volatility
could incorporate all the effects captured by local volatility and
a range of other empirical effects e.g. greater variability in
observed volatility. However the trade-off associated with
improved volatility modelling has been at the expense  of
analytical tractability.

\newpage
\bibliographystyle{plain}
\addcontentsline{toc}{section}{References}
\bibliography{Ref}

\end{document}